\documentclass[12pt,a4paper]{article}
\usepackage{amsmath,amsfonts,epsfig}
\usepackage{amssymb}
\usepackage{mathrsfs}
\usepackage{hyperref}
\newcommand{\beq}{\begin{eqnarray}}
\newcommand{\eeq}{\end{eqnarray}}

\newcommand{\be}{\begin{equation}}
\newcommand{\ee}{\end{equation}}

\newcommand{\bm}{\begin{multline}}
\newcommand{\fm}{\end{multline}}

\begin{document}
\numberwithin{equation}{section}
\setlength{\unitlength}{.8mm}

\begin{titlepage} 
\vspace*{0.5cm}
\begin{center}
{\Large\bf Extensive numerical study of a D-brane, anti-D-brane system in $AdS_5/CFT_4$}
\end{center}
\vspace{1.5cm}
\begin{center}
{\large \'Arp\'ad Heged\H us}
\end{center}
\bigskip

\vspace{0.1cm}

\begin{center}
Wigner Research Centre for Physics,\\
H-1525 Budapest 114, P.O.B. 49, Hungary\\ 
\end{center}
\vspace{2.5cm}
\begin{abstract}
In this paper the hybrid-NLIE approach of \cite{BH1} is extended to the ground state of 
a D-brane anti-D-brane system in AdS/CFT. The hybrid-NLIE equations presented in the paper
are finite component alternatives of the previously proposed TBA equations and they
admit an appropriate framework for the numerical investigation of the ground state of the problem.
Straightforward numerical iterative methods fail to converge, thus new numerical methods
are worked out to solve the equations. Our numerical data confirm the previous TBA data.
In view of the numerical results the mysterious $L=1$ case is also commented in the paper.
\end{abstract}

\end{titlepage}

\section{Introduction}

In this paper in the context of AdS/CFT \cite{adscft1,adscft2,adscft3} we study numerically the ground state
of a pair of open strings stretching between two coincident $D3$-branes with opposite
orientations in $S^5$ of $AdS_5 \times S^5$.
The main motivation for the study is that according to string-theory the ground state 
of such a configuration is expected to be tachyonic for large values of the 't Hooft coupling \cite{DH}.
In our work we rely on the perturbatively discovered and later "all loop conjectured" 
 integrability  \cite{Over} of both the $AdS_5 \times S^5$ super-string and the dual large $N$ gauge theory. 
 For string configurations with D-branes integrability enabled one to describe string configurations
ending on different types of D-branes as 1-dimensional integrable scattering theories with boundaries
\cite{DeWolfe:2004zt,Berenstein:2005vf,Hofman:2007xp,Correa:2008av,CRY}.
This formulation of the problem makes it possible to go beyond the approaches of perturbative
gauge and string theories being valid for small and large values of the 't Hooft coupling respectively,
and to determine the exact spectrum of the model at any value of the coupling constant. However, even with the help
of the powerful techniques offered by integrability, the exact analytical solution of the problem is not possible.
Remarkeble analytical results are available in the small \cite{Bajnok:2012bz,Leurent:2012ab,Leurent:2013mr,Volinuj,Marboe:2014sya,Bajnok:2013sya} 
and large \cite{Gromov:2009tq,Gromov:2010vb,Gromov:2014bva,Bajnok:2013sza} coupling regimes, but the determination
of the spectrum at any value of the coupling constant can only be carried out by high precision numerical solution
\cite{Gromov:2009zb,Frolov:2010wt,Frolov:2012zv} of the corresponding nonlinear integral equations.

In our paper we consider the case, when the two $D3$-branes are giant gravitons \cite{McGreevy:2000cw}, 
namely they carry $N$ units of angular momenta in $S^5$. If the $S^5$ of $AdS_5 \times S^5$ is parametrized
by three complex coordinates $X,Y,Z$ satisfying the constraint: $|X|^2+|Y|^2+|Z|^2=1$, then our 
$D3$-brane and anti-$D3$-brane are given by the conditions $Y=0$ and $\bar{Y}=0$ respectively. They
wrap the same $S^3$, but with opposite orientation. As a consequence of Gauss law such a system can
support only even number of open strings. For this reason we study the minimal number of allowed
open strings, a single pair, ending on our D-brane anti-D-brane ($D\bar{D}$) system with open string
angular momenta $L$ and $L'$.

On the large $N$ gauge theory side a $Y=0$ brane is represented by a determinant
 operator \cite{Balasubramanian:2001nh} composed of $N$ copies of the field $Y$:
\begin{equation}
{\cal O}_Y=\det Y=\epsilon^{a_1\cdots a_N}_{b_1\cdots b_N}
Y_{a_1}^{b_1}\cdots Y_{a_N}^{b_N} \label{detYY}
\end{equation}
where $a_i$ and $b_i$ are color indices and $\epsilon$ is a product of two 
regular epsilon tensors
$\epsilon^{a_1\cdots a_N}_{b_1\cdots b_N}=
\epsilon^{a_1\cdots a_N}\epsilon_{b_1\cdots b_N}$.
The local operator corresponding to an open string ending on a $Y=0$ giant graviton
can be obtained from (\ref{detYY}) by replacing one $Y$ field with an adjoint valued operator
${\cal W}$ \cite{DH33}:
\begin{equation}
{\cal O}_Y^{\cal W}=\epsilon^{a_1\cdots a_N}_{b_1\cdots b_N}
Y_{a_1}^{b_1}\cdots Y_{a_{N-1}}^{b_{N-1}}{\cal W}^{b_N}_{\ a_N}\,.
\end{equation}
The gauge theory description of a pair of open strings stretching between two $D$-branes is given by
a double determinant operator, such that the string insertions ${\cal W}$ and ${\cal V}$ connect the
two determinants of the $Y$ fields\footnote{The ground state of such string states is BPS.}:
\begin{equation}
\label{detYYWV}
{\cal O}_{Y,Y}^{{\cal W}, {\cal V}}=
\epsilon^{a_1\cdots a_N}_{b_1\cdots b_N}
Y_{a_1}^{b_1}\cdots Y_{a_{N-1}}^{b_{N-1}}
\,\epsilon^{c_1\cdots c_N}_{d_1\cdots d_N}
Y_{c_1}^{d_1}\cdots Y_{c_{N-1}}^{d_{N-1}}
{\cal W}^{d_N}_{\ a_N}{\cal V}^{b_N}_{\ c_N} 
\end{equation}
  Unfortunately, the precise gauge theory dual of the $D\bar{D}$-system of our interest is not known.
In \cite{DH} it was approximated by a double determinant operator similar to (\ref{detYYWV}), 
but in one of the determinants the $Y$ fields are replaced with $\bar{Y}$ fields\footnote{According to
the argument of \cite{DH} the correct state might have other structures involving the fields $Y$ and $\bar{Y}$, but
should be similar to the double determinant form (\ref{YYbar}) and the mixing with other fields seem to be
suppressed at large $N$}:
\begin{equation}
\label{YYbar}
{\cal O}_{Y\bar Y}^{{\cal W}, {\cal V}}=
\epsilon^{a_1\cdots a_N}_{b_1\cdots b_N}
Y_{a_1}^{b_1}\cdots Y_{a_{N-1}}^{b_{N-1}}
\,\epsilon^{c_1\cdots c_N}_{d_1\cdots d_N}
\bar Y_{c_1}^{d_1}\cdots\bar Y_{c_{N-1}}^{d_{N-1}}
{\cal W}^{d_N}_{\ a_N}{\cal V}^{b_N}_{\,c_N}\,.
\end{equation}
For the ground state the insertions are ${\cal W}=Z^L$ and ${\cal V}=Z^{L'}$ respectively.
Based on one-loop results  the planar dilatation operator is expected to act independently on
the two words ${\cal W},{\cal V}$ corresponding to the open string states \cite{DH}:
\begin{equation}
\Delta [{\cal O}_{Y\bar Y}^{{\cal W}, {\cal V}}] = \Delta_{\rm bare} [{\cal O}_{Y\bar Y}^{{\cal W}, {\cal V}}] 
+ \delta \Delta [{\cal W}_{Y\bar Y}] + \delta \Delta [{\cal V}_{\bar Y Y}].
\label{dim split}
\end{equation} 
This observation allows us to apply the boundary Thermodynamic Bethe Ansatz technique (BTBA) \cite{LeClair:1995uf} 
to each open string separately. The necessary ingredients of this technique are the boundary reflection
factors \cite{Hofman:2007xp,Correa:2009mz,Palla:2011eu,Ahn:2010xa} and the asymptotic Bethe equations 
of the problem \cite{DH}.
Unfortunately, apart from some very special cases \cite{Dru-int-WL,CMS,Bajnok:2013sya}, it is still unknown how to derive BTBA 
equations for a general non-diagonal scattering theory in the context of the thermodynamical considerations of \cite{LeClair:1995uf}.
This is why in \cite{DH} the $Y$-system \cite{GKV09,Bombardelli:2009ns,Arutyunov:2009ur} and the related discontinuity \cite{Cavaglia:2010nm} equations
supplemented by analyticity assumptions compatible with the asymptotic solution \cite{Balog:2011nm} were used to
derive BTBA equations for the nonperturbative study of the ground state of the $D\bar{D}$-system \cite{DH}. 

The BTBA description of the system is an infinite set of nonlinear integral equations. The numerical 
solution of the equations \cite{DH} showed that the ground state energy is a monotonously decreasing 
function of the coupling constant\footnote{Throughout the paper the relation between $g$ and the 't Hooft
coupling $\lambda$ is given by: $\lambda=4 \pi^2 g^2$.} $g$.
 The analytical investigation of the large rapidity and large
index behavior of the $Y$-functions of the BTBA revealed that the usual BTBA description of the system
breaks down when the energy of an open string state with angular momentum $L$ gets close to the critical value:
$E_c(L)=1-L$. This point was interpreted in \cite{DH} as a transition point where the ground state becomes
tachyonic.
Approaching the critical point the contribution of
infinitely many $Y$-functions must be taken into account to get accurate numerical result for the 
energy\footnote{This means that the usual truncation procedure for solving the infinite set of TBA equations
is not applicable to such a system}.
This fact suggests reformulating the finite size problem in terms of finite number of unknown functions.
The possible candidates could be the FiNLIE \cite{Gromov:2011cx}, the quantum spectral curve (QSC) \cite{Gromov:2013pga,Gromov:2014caa}
or the hybrid-NLIE (HNLIE) \cite{BH1} formulation of the problem. 
Since at present it is not known (not even for the Konishi problem) how to use the
analytically very efficient \cite{Gromov:2014bva,Volinuj}
QSC method for numerical purposes, we choose the HNLIE method to reformulate the finite size problem of the $D\bar{D}$-system.
In this paper we transformed the infinite set of boundary TBA equations \cite{DH} into a finite set of hybrid-NLIE
type of  nonlinear integral equations. We perform the extensive numerical study of these type of equations in order to
get as close to the special $E_{BTBA}=1-L$ critical point as it is possible. 

 Our numerical results
reproduce the numerical evaluation of the boundary L\"uscher formula \cite{Correa:2009mz,Bajnok:2010ui} 
in the linear approximation, and the numerical BTBA results of \cite{DH} as well. 
These numerical comparisons give further numerical checks
on the hybrid-NLIE technique of \cite{BH1}. Unfortunately, as $g$ increases new local singularities 
enter the HNLIE formulation of the problem. Thus we could not approach very close to the critical point.
Nevertheless, in the range of $g$ where physically acceptable numerical results were obtained, the HNLIE
results could give higher numerical precision than that of the BTBA and also some interesting facts could be
read off from our numerical data.

During the numerical solution of the HNLIE equations straightforward numerical iterative 
methods failed to converge, thus new numerical methods were worked out to solve the equations.

The ground state of the $L=1$ state is a very special case, since there the critical point is right
at $g=0$ and so far neither perturbative field theory computations nor the boundary L\"uscher formula
could provide a finite quantitative answer to the anomalous dimension of this state. 
On the integrability side the HNLIE approach allows us to get some numerical insight into this problem. 

The outline of the paper is as follows:
Section 2. contains the HNLIE equations. In section 3. the numerical method is described.
In section 4. the numerical results and their interpretation is presented.
Section 5. contains some comments on the mysterious $L=1$ case and finally our conclusion is given
in section 6.
Various notations, kernels of the integral equations together with the necessary asymptotic solutions
are placed in the appendices of the paper.

\section{The HNLIE equations}

In this section we transform the previously proposed BTBA equations of \cite{DH} for the ground 
state of our D-brane anti-D-brane system to finite component hybrid-NLIE equations.
For presentational purposes we group the equations into 3 types.
There are TBA-type equations, horizontal $SU(2)$ hybrid-NLIE type equations, and 
vertical $SU(4)$ hybrid-NLIE type equations. They together form a closed set of
nonlinear integral-equations, which are solved numerically in this paper.
As it is usual, structurally the equations consist of source terms plus 
convolutions containing coupling dependent kernels and nonlinear combinations of the 
unknown functions. The objects appearing in the arguments of the source functions
are subjected to quantization conditions, but similarly to the boundary TBA description \cite{DH}, 
due to the $u \rightarrow -u$ symmetry of the
problem they are tied to the origin of the complex plane, thus extra quantization conditions
are unnecessary to be imposed, since they are automatically satisfied by symmetry.
Since these source term objects have fixed positions their positions are exactly the same as
that of their asymptotic counterparts. This fact saves us from the tedious computation
of the source terms, since if we take the difference of the exact equations and their
asymptotic counterparts the source terms cancel from the equations. To be pragmatic and save time 
and space, the equations will be presented in such a difference form.
Thus for any combination $f$ of the unknown functions, we introduce the notation 
$\delta f(u)=f(u)-f^o(u)$, where $f^o(u)$ is the asymptotic counterpart of $f$.
Having introduced this notation, we start the presentation of the equations by the TBA-type
part. For the labeling of the $Y$-functions we use the string-hypothesis \cite{AFshyp} based notations
of \cite{AFS}. For the presentation of the equations a few more notations need to be introduced:
\begin{equation}
{\mathscr L}_{\pm}=\log\left[\tau^2 \left( 1-\frac{1}{Y_{\pm}}\right) \right], \qquad 
{\mathscr L}_{m}=\log\left[\tau^2 \left( 1+\frac{1}{Y_{m|vw}}\right) \right], \qquad 
\tau(u)=\tanh(\frac{\pi g u}{4}).
\end{equation}
For later numerical purposes we re-parametrize $\log Y_Q$ by the formula:
\begin{equation}
\log Y_Q(u)=-2 L \, \log \frac{x^{[Q]}(u)}{x^{[-Q]}(u)}+\log \bar{y}_Q(u) +c_Q+\varepsilon \, \log\left(u^2+\frac{(Q+1)^2}{g^2}\right), \qquad Q=1,2,... \label{separ}
\end{equation}
such that $c_Q$ is the constant value of $\log Y_Q$ at infinity and $\varepsilon$ is minus twice the energy\footnote{The $\log$ 
multiplier of $\varepsilon$ in (\ref{separ}) is chosen not to modify the
constant term in the
large $u$ behavior and to satisfy $\frac{Y_Q^{+} \, Y_Q^{-}}{Y_{Q-1}\, Y_{Q+1}}
\frac{Y_{Q-1}^o\, Y_{Q+1}^o}{Y_Q^{o+} \, Y_Q^{o-}}
=\frac{\bar{y}_Q^{+} \, \bar{y}_Q^{-}}{\bar{y}_{Q-1}\, \bar{y}_{Q+1}} \frac{\bar{y}^o_{Q-1}\, \bar{y}^o_{Q+1}}{\bar{y}_Q^{o+} \, \bar{y}_Q^{o-}}$
, which is the LHS of an important $Y$-system equation divided by its asymptotic counterpart.}:
$\varepsilon=-2 \, E_{BTBA}$. From the TBA equations of the problem \cite{DH}, it follows 
that $\delta c_Q=c_Q-c_Q^o\equiv \delta c$ is $Q$-independent, and
for small $g$, $\log \bar{y}_Q$ is a smooth deformation of its asymptotic counterpart, such that
$\delta\log \bar{y}_Q$ tends to zero at infinity.\footnote{$\log Y_Q$ cannot be
considered as smooth deformation of $\log Y_Q^o$, because 
$\log Y_Q-\log Y_Q^o \sim \varepsilon \log |u|$ diverges for large $u$ at any $g$.
On the other hand $\log \bar{y}_Q-\log \bar{y}_Q^o$ is small for any $u$ at small $g$ 
and tends to zero at infinity.} 

Using this decomposition the following notations are need to be introduced:
\begin{equation}
L_Q=\log(1+Y_Q), \qquad \delta\!R_Q=\log(1+Y_Q)-\delta\log\bar{y}_Q.
\end{equation}
Then the TBA-type equations take the form:
\begin{equation}
\delta \! \log Y_{m\vert vw}=
\delta\log\left[(1+Y_{m+1\vert vw})(1+Y_{m-1\vert vw})\right]\star s-
 \log (1+Y_{m+1})\star s,\quad 2\leq m \leq p_0-2,
\label{TBAmvw}
\end{equation}
\begin{equation}
\delta\log Y_{1\vert vw}=
\delta\log\left(1+Y_{2\vert vw}\right)\star s-\log(1+Y_2)\star s
+\delta\log\left[\frac{1-Y_-}{1-Y_+}\right]\ \hat\star\ s,\label{TBA1vw}
\end{equation}
\begin{eqnarray}
\delta \! \log \bar{y}_Q&=& 2\, \delta{\mathscr L}_{Q-1} \star s-(\delta R_{Q-1}+\delta R_{Q+1} \star s)
,\quad Q\geq2,\label{TBAQ}\\
\delta\log\frac{Y_-}{Y_+}&=&
-\sum_{Q=1}^{p_0-2} \log (1+Y_Q)\star K_{Qy}-\Omega(K_{Qy}).
\label{YmperYp} \\
\delta \! \log(Y_+Y_-)&=& 
2 \delta \! \log\left[\frac{1+Y_{1\vert vw}}{1+Y_{1\vert w}}
\right]\star \! s+\! \! \sum_{Q=1}^{p_0-2} \log (1+Y_Q)\star\left[
-K_Q+2K^{Q1}_{xv}\star \! s\right] \nonumber \\
&-&\Omega(K_Q)+2 \, \Omega(K^{Q1}_{xv}\star \! s) \label{YpYm}
\end{eqnarray}
For  $Y_1$ the modified hybrid form \cite{BH2} of the BTBA equations is used,
\begin{equation}
\begin{split}
\delta \! \log \bar{y}_1=& 
2\delta\log(1+Y_{1|vw})\star s\ \hat\star\ K_{y1}\\
&-2\delta\log\left[\frac{1-Y_-}{1-Y_+}\right]\ \hat\star\ s\star K^{11}_{vwx}
+2\delta{\mathscr L}_-\ \hat\star\ K^{y1}_-+2\delta{\mathscr L}_+\ \hat\star\ K^{y1}_+\\
&+\sum_{Q=1}^{p_0-2} \log(1+Y_{Q})\star\left[K^{Q1}_{{\mathfrak{sl}(2)}}
+2s\star K^{Q-1,\,1}_{vwx}
\right] \\
&+\Omega(K^{Q1}_{{\mathfrak{sl}(2)}})+2 \Omega(s\star K^{Q-1,\,1}_{vwx}),
\end{split}
\label{hybrid}
\end{equation}
where $p_0$ is the index limit starting from which the upper part of the TBA equations
is replaced by an $SU(4)$ NLIE of \cite{BH1} (See figure \ref{YHNLIE}.).
For any kernel vector appearing in the TBA equations $\Omega({\cal K}_Q)$ denotes the residual sum 
$\sum\limits_{Q=p_0-1}^{\infty} L_Q \star {\cal K}_Q$, and following the
method of \cite{BH2} for $p_0\geq 4$ it can be expressed by next to nearest neighbor $Y$-functions as follows:
\begin{eqnarray}
\Omega({\cal K}_Q)&=&\delta R_{p_0-1} \star \sigma_{\frac12} \star {\cal K}_{p_0-2}-\delta R_{p_0-2} \star \sigma_{\frac12} \star
{\cal K}_{p_0-1} \nonumber \\
&+&2 \delta r_{p_0-2} \star s_{\frac12} \star {\cal K}_{p_0-2}- 2 \delta r_{p_0-3} \star s_{\frac12} \star {\cal K}_{p_0-1}, \label{OmegaKQ}
\end{eqnarray}
where $r_m=\log(1+Y_{m|vw})$, the kernels $s,s_{\frac12},\sigma_{\frac12}$ are hyperbolic functions \cite{BH2},
\begin{equation}
s(u)=\frac{g}{4 \, \cosh \frac{\pi \, g \, u}{2}}, \qquad
s_{\frac12}(u)=\frac12 \, s(\frac{u}{2}),\qquad
\sigma_{1/2}(u)=\frac{g}{2\sqrt{2}}\,\frac{\cosh\frac{\pi gu}{4}}
{\cosh\frac{\pi gu}{2}},
\end{equation}
while the other TBA kernels can be found in appendix A.
As a consequence of the re-parametrization (\ref{separ}) the two constants $\delta c$ and $\varepsilon$ 
also become part of the set of equations\footnote{Here the $\star$ notation means simply integration from $-\infty$ to $\infty$.}:
\begin{equation}
\varepsilon=\frac{1}{2 \pi}\sum_{Q=1}^{p_0-2} \, L_Q \star \frac{d\tilde{p}^Q}{du}+\frac{1}{2 \pi}\Omega(\frac{d\tilde{p}^Q}{du}), \label{eps}
\end{equation}
\begin{equation}
\begin{split}
\delta c=& 
2\delta\log(1+Y_{1|vw})\star s\ \hat\star\ CK_{y1}\\
&-2\delta\log\left[\frac{1-Y_-}{1-Y_+}\right]\ \hat\star\ s\star CK^{11}_{vwx}
+2\delta{\mathscr L}_-\ \hat\star\ CK^{y1}_-+2\delta{\mathscr L}_+\ \hat\star\ CK^{y1}_+\\
&+\sum_{Q=1}^{p_0-2} \log(1+Y_{Q})\star\left[CK^{Q1}_{{\mathfrak{sl}(2)}}
+2s\star CK^{Q-1,\,1}_{vwx}
\right] \\
&+\Omega(CK^{Q1}_{{\mathfrak{sl}(2)}})+2 \Omega(s\star CK^{Q-1,\,1}_{vwx}),
\end{split}
\label{chybrid}
\end{equation}
where for any kernel ${\cal K}$: $C{\cal K}(u)$ denotes the constant term in the large $v$ expansion
of ${\cal K}(u,v)$.\footnote{Here we note that only the dressing kernel has logarithmically divergent
term in its large $v$ expansion, all the other kernels has either constant term or they simply vanish at infinity.}
As we mentioned $-\varepsilon/2$ is the TBA energy, thus (\ref{eps}) gives the energy formula in our
formulation of the finite size problem. The asymptotic forms of the $Y$-functions necessary for the formulation
of (\ref{TBAmvw}-\ref{chybrid})
are listed in appendix D.
To close the discussion of the TBA-type equations we note that equations (\ref{YmperYp}) and (\ref{YpYm}) determine
$Y_{\pm}$ up to an overall sign factor.
The sign factor can be fixed from the asymptotic solution and its value is $-1$.
Thus the fermionic Y-functions can be expressed in terms of the LHS of (\ref{YmperYp}) and
(\ref{YpYm}) by the formula:
\begin{equation}
Y_{\mp}=-e^{\frac12 \log Y_{+} Y_{-}\pm \frac12 \log \frac{Y_{-}}{Y_{+}}}.
\end{equation}
The horizontal $SU(2)$ wing of the TBA is resumed by an $SU(2)$-type NLIE \cite{RyoNLIE,BH1}, which
in our case takes the form:
\begin{equation}
\delta \! \log (-b)\!=\!s^{[1-\gamma]}\star\delta\log(1+Y_{1|w})+G\star\delta\log(-1-b)-
G^{[-2\gamma]} \star \delta\log(-1-\bar{b}),
\end{equation}
\begin{equation}
\delta\log(-\bar{b})\!=\!s^{[\gamma-1]}\star\delta\log(1+Y_{1|w})+G\star\delta\log(-1-\bar{b})-
G^{[2\gamma]}\star\delta\log(-1-b),
\end{equation}
\begin{equation}
\delta\!\log Y_{1\vert w}=
s^{[\gamma-1]} \star \delta\!\log(-1-b)
+s^{[1-\gamma]} \star \delta\!\log(-1-\bar{b})+ 
\delta\!\log\left[\frac{1-\frac{1}{Y_-}}{1-\frac{1}{Y_+}}\right]
\ \hat\star\ s \,,\label{TBA1w}
\end{equation}
where $0<\gamma<1/2$ is a contour shift parameter, the kernel $G$ is given by (\ref{G}) and the 
asymptotic solution for $b$ and $\bar{b}$ is given in appendix D\footnote{In practice $b$ and 
$\bar{b}$ are complex conjugate of each other.}.
The upper $SU(4)$ NLIE of \cite{BH1} is attached to the TBA equations at the $p_0$-th node.
The upper NLIE is for 12 complex unknown functions: $b_A$ and $d_A, \quad A=1,...,6$. They are
combinations of the $T$-functions of the upper wing $SU(4)$ B\"acklund-hierarchy \cite{BH1}. Their relations
to the unknowns introduced in \cite{BH1} are given by (\ref{buj},\ref{duj}) in appendix B and their asymptotic
 forms  are given in appendix C. Using the notation $B_A=1+b_A$ and $D_A=1+d_A$, the equations they satisfy 
take the form:
\begin{equation}
\delta \! \log b_A=\sum_{A'} (G_{bB})_{AA'} \star \delta \! \log B_{A'}+
\sum_{A'} (G_{bD})_{AA'} \star \delta \! \log D_{A'}+E_A, \label{bA}
\end{equation}
\begin{equation}
\delta \! \log d_A=\sum_{A'} (G_{dB})_{AA'} \star \delta \! \log B_{A'}+
\sum_{A'} (G_{dD})_{AA'} \star \delta \! \log D_{A'}+\bar{E}_A, \label{dA}
\end{equation}
where the kernels are given in (\ref{93a}-\ref{KdD}). The shifts in the kernels which
is equivalent to fixing the lines on which the NLIE variables live, are chosen
in a symmetrical way and fixed as follows:
\begin{equation}
\underline{\gamma}=\{\gamma_a\}=\{\gamma^{(3)}_1,\gamma^{(3)}_2,\gamma^{(3)}_3,\gamma^{(2)}_1,\gamma^{(2)}_2,\gamma^{(1)}_1\}=\frac{1}{12} (-9,-1,5,-3,3,1),
\label{vgamma}
\end{equation}
\begin{equation}
\underline{\eta}=\{\eta_a\}=\{\eta^{(3)}_1,\eta^{(3)}_2,\eta^{(3)}_3,\eta^{(2)}_1,\eta^{(2)}_2,\eta^{(1)}_1\}= 
\frac{1}{12} (-5,1,9,-3,3,-1).
\label{veta}
\end{equation}
This choice satisfies the constraint inequalities of \cite{BH1} and satisfy the relation \newline
$\underline{\gamma}=-{\mathfrak M} \, \underline{\eta}$ with ${\mathfrak M}$ given by (\ref{M}).
Its advantage is that choosing the $C=0$ asymptotic solution from appendix C to formulate the equations,
the $b$- and $d$-type variables are related in a simple manner:
\begin{equation}
b(-u)={\mathfrak M} d(u). \label{bmMd}
\end{equation}  
In practice this reduces to half the number of $SU(4)$ NLIE variables.
The vectors $E_A$ and $\bar{E}_A$ are
conjugate to each other and they give the TBA input into the upper NLIE.
To give their form we introduce the notations: 
\begin{equation}
{\eta_1 }=Y_{p0-1|vw}^{[\epsilon_1]}, \qquad {\bar{\eta}_1 }=Y_{p0-1|vw}^{[\epsilon_3]}, \qquad \epsilon_1=-\epsilon_3=-\frac{7}{12}, 
\end{equation}
\begin{equation}
E_1=s^{[\frac56]} \star \delta \! \log (1+\eta_1),\qquad \qquad
E_3=s^{[\frac56]} \star \delta \! \log (1+\bar{\eta}_1), \qquad \qquad
E_5=E_6=0,
\end{equation}
\begin{equation}
E_2=\frac12 s^{[\frac12]} \star \delta \! \log (1+\eta_1)-\frac12 s^{[\frac23]} \star \delta \! \log 
(1+\bar{\eta}_1)+\varepsilon_2 +i \, \varphi_2,
\end{equation}
\begin{equation}
E_4=-\frac12 s^{[\frac13]} \star \delta \! \log (1+\eta_1)-\frac12 s^{[\frac56]} \star \delta \! \log 
(1+\bar{\eta}_1)+\varepsilon_4 +i \, \varphi_4,
\end{equation}
where
\begin{equation}
\varepsilon_2(u)=\frac{i}{2 \pi} \int\limits_{0}^{\infty} dv \,\, \delta\! R_{p_0}(v) \,
\left\{  \frac{1}{u-v-\frac{i}{12 g}}+\frac{1}{u+v-\frac{i}{12 g}}  \right\},
\end{equation}
\begin{equation}
\varepsilon_4(u)=\frac{i}{2 \pi} \int\limits_{0}^{\infty} dv \,\, \delta\! R_{p_0}(v) \,
\left\{  \frac{1}{u-v-\frac{i}{4 g}}+\frac{1}{u+v-\frac{i}{4 g}}  \right\},
\end{equation}
\begin{equation}
\varphi_2(u)=\int\limits_{-\infty}^{\infty} dv \,\, \delta\! \log (1+\eta_1(v)) \,
\left\{  \varphi(u-v+\frac{i}{2 g}) +\varphi(u+v-\frac{2 \,i}{3 g}) \right\},
\end{equation}
\begin{equation}
\varphi_4(u)=\int\limits_{-\infty}^{\infty} dv \,\, \delta\! \log (1+\eta_1(v)) \,
\left\{  \varphi(u-v+\frac{i}{3 g}) +\varphi(u+v-\frac{5 \,i}{6 g}) \right\},
\end{equation}
with 
\begin{equation}
\varphi(u)=\frac{g}{8 \,\pi} \left\{i\, \psi(\frac{1}{4}-\frac{i\, u\, g}{4})-
i\, \psi(\frac{1}{4}+\frac{i\, u\, g}{4})-\pi\, \tanh(\frac{\pi\, g\, u}{2}) \right\}.
\end{equation}
The last set of equations gives, how the upper NLIE variables couple to the TBA
part of the equations.
\begin{equation}
\delta\!\log Y_{p_0-2|vw}=s^{[-\epsilon_1]} \star \delta\!\log(1+\eta_1)+
s \star \delta\!\log (1+Y_{p_0-3|vw})-s \star L_{p_0-1}, \label{Yvwp0m2}
\end{equation}
\begin{equation}
\begin{split}\delta\!\log \eta_1&=s^{[-1+\epsilon_1-\gamma_1]} \star \delta\!\log B_1+
s^{[1+\epsilon_1-\eta_1]} \star \delta\!\log D_1-s^{[\epsilon_1-\gamma_2]} \star
\delta\!\log B_2 \\
&+s^{[\epsilon_1]} \star \delta\!\log (1+Y_{p_0-2|vw})
-s^{[\epsilon_1]} \star L_{p_0},
\end{split}
\end{equation}
\begin{equation}
\begin{split}
\delta\!\log \bar{\eta}_1&=s^{[-1+\epsilon_3-\gamma_3]} \star \delta\!\log B_3+
s^{[1+\epsilon_3-\eta_3]} \star \delta\!\log D_3-s^{[\epsilon_3-\eta_2]} \star
\delta\!\log D_2  \\
&+s^{[\epsilon_3]} \star \delta\!\log (1+Y_{p_0-2|vw})
-s^{[\epsilon_3]} \star L_{p_0},
\end{split}
\end{equation}
\begin{equation}
\begin{split}
\delta\!\log \bar{y}_{p_0}= s^{[-1-\gamma_2]} \star 
\left[ \delta\!\log\bar{b}_2-\delta\!\log B_2 \right]+
s^{[1-\eta_2]} \star \left[ \delta\!\log\bar{d}_2-\delta\!\log D_2 \right] \\
+s^{[-\gamma_3]} \star \delta\!\log \frac{B_3}{b_3}+
s^{[-\eta_1]} \star \delta\!\log \frac{D_1}{d_1} 
+s^{[-\epsilon_1]} \star \delta\!\log \frac{1+\eta_1}{\eta_1} \\
+s^{[-\epsilon_3]} \star \delta\!\log \frac{1+\bar{\eta}_1}{\bar{\eta}_1}-
s \star \delta\! R_{p_0-1}, \label{lasteq}
\end{split}
\end{equation}
where $\bar{b}_2$ and $\bar{d}_2$ are from the re-parametrization of
$b_2$ and $d_2$:
\begin{equation}
b_2(u)=\eta \, \left(\frac{1}{x_s^{[-p_0+\gamma_2]}(u)}\right)^{2L} \, \exp\left\{ 
\varepsilon \left[ \log\left(u+i\,\frac{\gamma_2-p_0-1}{g}\right)
+i \,\frac{\pi}{2} \right]+\frac{\delta c}{2}
\right\} \, \bar{b}_2(u), \label{b2bar}
\end{equation}
\begin{equation}
d_2(u)=\eta \, \left(\frac{1}{x_s^{[p_0+\eta_2]}(u)}\right)^{2L} \, \exp\left\{ 
\varepsilon \left[ \log\left(u+i \, \frac{\eta_2+p_0+1}{g}\right)-i \,\frac{\pi}{2} \right]+\frac{\delta c}{2}
\right\} \, \bar{d}_2(u), \label{d2bar}
\end{equation}
with $\eta=\pm 1$ being a global sign factor. 
Similarly to the definition of $\bar{y}_Q$, also here the benefit of using $\bar{b}_2$ and $\bar{d}_2$
is that, for small $g$, $\log\bar{b}_2$ and $\log\bar{d}_2$ are smooth deformations of their asymptotic counterparts, 
and in addition $\delta\!\log \bar{b}_2$ and $\delta\!\log \bar{d}_2$ vanishes at infinity, which is necessary
for the convergence of certain integrals.
 The decompositions (\ref{b2bar}),(\ref{d2bar}) are chosen to be compatible
with the functional relation $b_2^{[-\gamma_2]} \, d_2^{[-\eta_2]}=Y_{p_0}$ in ref. \cite{BH1}.
Equations (\ref{TBAmvw})-(\ref{lasteq}) constitute our complete set of nonlinear integral equations, which governs the 
finite size dependence of the vacuum of our D-brane anti-D-brane system.

\section{The numerical method}

Here we describe our numerical method for solving the hybrid-NLIE equations presented in the previous section.
During the iterative numerical solution of the equations we faced with very serious convergence problems, which
forced us to work out such a method that overcomes all the difficulties emerged. Our numerical method can be applied
to solve other type of nonlinear integral equations as well. The power of the method is shown by the fact that
numerical convergence was reached even in such cases, when the solution was physically unacceptable.
\newline
The numerical method consist of two main steps, namely:
\begin{itemize}
\item Discretization of the equations
\item Iterative solution.
\end{itemize}
The first step involves the discretization of the unknown functions and kernels, furthermore
the discrete approximate representation of the convolutions. Having carried out the appropriate
discretization of the problem, the equations are considered as large nonlinear algebraic set of
equations. Thus eventually instead of integral equations we solve discrete algebraic equations.
In this paper we will present two methods to solve them numerically.

\subsection{Discretization of the problem}

The discretization serves two goals. First it allows us to reduce the numerical problem
from solving integral equations to solving algebraic equations. Second choosing the discretization
points appropriately it reduces the number of degrees of freedom as much as it is possible
to reach the desired numerical accuracy.
In our actual numerical computation instead of $u$ of section 2. we used the new rapidity
$u \rightarrow \frac{u}{g}$, because with such a scaling almost all the rapidity difference
dependent kernels become $g$ independent. Thus for example $Y_{\pm}(u)$ will be defined in
$[-2 g ,2 g]$. To decrease the number of discretization points the $u \to -u$ symmetry
of the problem is exploited. This means that the $Y$-functions are to be discretized only
on $[0,\infty]$ or $[0, 2 g]$ and as for the NLIE variables it is enough to discretize the $b$-
and $d$- type variables on $[0,\infty]$.
Since we do not want to introduce any cutoff in the rapidity space first we transform the $u\in[0, \infty]$
interval to a finite interval $t\in[0,B(a)]$ through the transformation formula:
\begin{equation}
u(t)=a\left(\frac{B(a)}{t}-1 \right), \qquad B(a)=2 \, \frac{g}{a}+1. \label{xt}
\end{equation}
This formula is chosen such that the branch point $2 g$ corresponds to $t=1$ for any choice of the
parameter $a$, where $a$ is a global scaling factor which changes from unknown to unknown.
We chose the values as follows:
for $Y_{1|w}$ $a=1$, for $b$ and $\bar{b}$ $a=2$, for $Y_{Q}$ and $Y_{Q-1|vw}$ $a=Q$, for $\eta_1$ and $\bar{\eta}_1$ $a=p_0$,
and finally for the $b$- and $d$-type NLIE functions $a=p_0$.
These values are chosen to preserve the smoothness\footnote{In our terms the lack of smoothness would not mean discontinuity, but the presence of 
rapidly changing parts and peaks.} of the transformed functions in the finite interval. 
After this transformation all of our unknown functions live on a finite interval. To discretize them
we used piecewise Chebyshev approximation. This means that we divide the finite interval into
subintervals and on each subinterval the functions are approximated by a given order Chebyshev series.
The choice of subintervals is not equidistant. The subintervals are placed more densely around the branch points,
since the function $x(u/g)$, which governs the decay of the massive $Y_Q$-functions,
has the largest change around this point. The advantage of the Chebyshev approximation is that if the function
is smooth enough on the subinterval, the coefficients of the Chebyshev series decay rapidly and the order of magnitude
of the last coefficient allows us to estimate the numerical errors of the procedure. Now we describe the discretization
method in more detail. Our functions are defined on either $[0,B(Q)]$ or on $[0,2 g]$. This is why two type of subinterval
vectors are defined $A_Q$ and $A_{\pm}$, such that the endpoints of the subintervals of $[0,B(Q)]$ are put into the vector $A_Q$ and
the endpoints of the subintervals of $[0,2 g]$ define $A_{\pm}$. Let $l_k$ be the order of the Chebyshev approximation, then using the general rules of the Chebyshev approximation,
a given function $f(t)$ is approximated in the $k$th subinterval $[A_{k-1},A_k]$ as:
\begin{equation}
f(t)\simeq \sum\limits_{j=1}^{l_k} \, c^{(k)}_j \, \hat{T}_{j-1}\left( \frac{t-\frac12 (A_k+A_{k-1})}{\frac12 (A_k-A_{k-1})}\right),
\qquad t \in [A_{k-1},A_k],
\label{Tkoz}
\end{equation}
where now the vector $A$ stand for either $A_Q$ or $A_{\pm}$, furthermore $\hat{T}_{j-1}$ are a slightly
modified Chebyshev polynomials\footnote{This slight modification
is only to write the approximation series (\ref{Tkoz}) in a more compact way.}
\begin{equation*}
\hat{T}_j(u)=\left\{ \begin{array}{llr}
T_j(u) & \mbox{if } & j\geq 1,\\
\frac12 & \mbox{if } & j=0,
\end{array}\right.
\end{equation*}
with $T_j(u)$ being the $j$th Chebyshev polynomial\footnote{The Chebyshev polinomials are defined by the formula: $T_j(u)\!=\!\cos(j \, \arccos u),\quad j=0,1,2... $}.
The coefficients $c^{(k)}_j$ are the Chebyshev coefficients of the function $f$, which can be computed from the 
sampling points of the Chebyshev approximation:
\begin{equation}
t^{(k)}_j=\frac12 (A_k-A_{k-1})\, c^{(i)}(l_k)+\frac12 (A_k+A_{k-1}), \qquad i=1, .., l_k
\end{equation}
by the simple formula:
\begin{equation}
c^{(k)}_j=\frac{2}{l_k} \, \sum\limits_{j_0=1}^{l_k} \, f(t^{(k)}_{l_k-j_0+1})\, \tilde{C}_{j_0,j},
\end{equation}
where $c^{(i)}(l_k)$ are the zeros of the $l_k$ order Chebyshev polynomial:
\begin{equation}
c^{(i)}(l_k)=-\cos \left[ \frac{\pi}{l_k}\left(i-\frac12\right)\right], \qquad \hat{T}_{l_k}(c^{(i)}(l_k))=0, \qquad i=1,...,l_k
\end{equation}
and $\tilde{C}_{k,i}$ is given by:
\begin{equation}
\tilde{C}_{k,i}=\cos \left[ \frac{\pi}{l_k}\left(k-\frac12\right)(i-1)\right], \qquad  \qquad i,k\in\{1,...,l_k\}.
\end{equation}
In our method the next step is to formulate the convolutions and the equations themselves in terms of the discrete values of our functions. Here 
will sketch the basic idea in some typical scenarios appearing in our equations. Then its application to the concrete
unknowns and kernels of the problem is straightforward.
If one takes the equations at the required discretized points $t^{(k)}_j$ the following typical pattern arises:
\begin{equation}
F(u(t^{(k')}_{j'}))\simeq \int\limits_{0}^{\infty} \, dv' \,L(v') \, K^S(v',u(t^{(k')}_{j'})) +\dots,
\end{equation}
where $K^S(u,v)=K(u,v)+K(-u,v)$ is the symmetrized kernel to exploit left-right symmetry of the problem for reducing to half
the number of variables. $F(u(t^{(k')}_{j'}))$ is intended to modelize the variables in the left-hand side of the equations taken 
at the discretized points of the transformed variable $t$ and $L(u)$ stands for some nonlinear combination of some unknown 
function of the equations\footnote{For example in the TBA-part $F(u)$ can be thought of as $\log Y(u)$ and $L(u)$ can be
$\log(1+Y(u))$ for any type of $Y$.}.
If $L(u(t))$ is discretized by a subinterval vector $A$ of $[0,B(a)]$,
then the numerical approximation of the right hand side goes as follows;
\begin{itemize}
\item First the integration variable is changed from $v'$ to $t$,
\item then on each subinterval $L(u(t))$ is approximated by its Chebyshev series,
\item finally the integration is carried out and the convolution is expressed in terms of the
discretized values of $L(u(t))$.
\end{itemize}
The final approximation formula takes the form:
\begin{equation}
\int\limits_{0}^{\infty} \, dv' \,L(v') \, K^S(v',u(t^{(k')}_{j'})) \simeq \sum\limits_{k=1}^{{\cal L}(A)}
\sum\limits_{j=1}^{l_k}\, L_{k,j} \, \left(\frac{2}{l_k} \sum\limits_{j_0=1}^{l_k} 
\tilde{C}_{l_k-j+1,j_0}\, {\cal K}^{k,j_0}_{k',j'}   \right), \label{intapproxt}
\end{equation}
where $L_{k,j}=L(u(t^{(k)}_j))$, ${\cal L}(A)$ denotes the dimension of $A$ and ${\cal K}^{k,j}_{k',j'}$ 
is the discretized convolution matrix given by the formula:
\begin{equation}
{\cal K}^{k,j}_{k',j'}=a \, B(a) \, \int\limits_{A_{k-1}}^{A_k} \, \frac{dt}{t^2} \,
\hat{T}_{j-1}\left( \frac{t-\frac12 (A_k+A_{k-1})}{\frac12 (A_k-A_{k-1})}\right) \,
K^S(u(t),u(t^{(k')}_{j'})). \label{tKSmatrix}
\end{equation}
In this manner a convolution is reduced to a discrete matrix-vector multiplication.

The other type of typical convolution is when the integration is taken from zero to $2 g$.
In certain cases the function $L(u)$ has square root behavior close to the branch points\footnote{
Such typical combinations are $\log\frac{1-Y_{-}}{1-Y_{+}}$ and $\log\frac{1-\frac{1}{Y_{-}}}{1-\frac{1}{Y_{+}}}$.}.
For such functions the truncated Chebyshev series does not give accurate approximation. In these cases not
the function $L(u)$ is approximated, but that part of it which remains after the elimination of the
square root behavior. Namely, we write $L(u)=\sqrt{4g^2-u^2} \, \hat{L}(u)$, then $\hat{L}(u)$ is approximated by 
a truncated Chebyshev series and finally the approximate discretized form of the corresponding convolution
is very similar to (\ref{intapproxt}):
\begin{equation}
\int\limits_{0}^{2 g} \, dv \,L(v)\,  K^S(v,u^{(k')}_{j'})\simeq \sum\limits_{k=1}^{{\cal L}(A_{\pm})}
\sum\limits_{j=1}^{l_k}\, \hat{L}_{k,j} \, \left(\frac{2}{l_k} \sum\limits_{j_0=1}^{l_k} 
\tilde{C}_{l_k-j+1,j_0}\, \hat{\cal K}^{k,j_0}_{k',j'}   \right), \label{intapproxu}
\end{equation}
where $\hat{L}_{k,j}=\hat{L}(v^{(k)}_j)$ and $\hat{\cal K}^{k,j}_{k',j'}$ 
is the square root factor modified version of (\ref{tKSmatrix});
\begin{equation}
\hat{\cal K}^{k,j}_{k',j'}= \int\limits_{A_{\pm,k-1}}^{A_{\pm,k}} dv \, \sqrt{4 g^2-v^2} \,
\, \hat{T}_{j-1}\! \left( \frac{v-\frac12 (A_{\pm,k}+A_{\pm,k-1})}{\frac12 (A_{\pm,k}-A_{\pm,k-1})}\right) \,
K^S(v,u^{(k')}_{j'}). \label{uKSmatrix}
\end{equation}
Here depending on the left hand side of the equation $u^{(k')}_{j'}$ can stand for $u(t^{(k')}_{j'})$,
$t\in[0,B(a)]$ for some $a$, or it can denote the sampling points on $[0,2 g]$.

Applying our discretization technique to all unknowns and convolutions of our equations, we can reduce
the integral equations to a discrete set of nonlinear algebraic equations. However, the transformation from integral equations
to algebraic equations is obviously not exact. The typical error comes from the fact that on each subinterval
the Chebyshev series is truncated, so the magnitude of the typical errors in our numerical method is governed by the
neglected terms of the Chebyshev series, which can be approximated by the magnitude of the last Chebysev coefficient.
In our case this is typically somewhere between $10^{-5}$ and $10^{-6}$.

The last step of our numerical method is the iterative solution starting from the asymptotic solution.

\subsection{The iterative solution}

Here we will describe two methods to solve our integral equations iteratively. Since our actual equations
have very complicated form, we will describe our methods using a model example, which has similar structure
to our equations.

Let the model equations take the form\footnote{For repeated indexes summation is understood.}:
\begin{equation}
\log y_a=f_a+G_{ab} \star \log(1+y_b), \label{modelleq}
\end{equation}
where $G_{ab}$ are some kernel matrices, $f_a$ are some source terms and $y_a$s are the unknown functions
of the problem. The solution of (\ref{modelleq}) is expanded around the asymptotic solution and the equations
are formulated in terms of the corrections.  
To fix the conventions, the correction functions $\delta y_a$ are defined by:
\begin{equation}
y_a=y^o_a\,(1+\delta y_a).
\end{equation}
As a consequence:
\begin{eqnarray}
\log y_a &=&\log y^o_a + \log(1+\delta y_a), \nonumber \\
\log (1+y_a) &=&\log (1+y^o_a) + \log(1+{\cal Y}_a \, \delta y_a), \qquad
 {\cal Y}_a=\frac{y^o_a}{1+y^o_a}.\nonumber
\end{eqnarray}
The source term is also expanded around its asymptotic counterpart:
$ f_a=f^o_a+\delta f_a$.
Then equations (\ref{modelleq}) can be reformulated in terms of the $\delta y_a$
functions as follows:
\begin{equation}
\log (1+\delta y_a)=\delta f_a+G_{ab} \star \log(1+ {\cal Y}_b \, \delta y_b). \label{dmodelleq}
\end{equation}
To define the iterative method, (\ref{dmodelleq}) are reformulated so that only
$O(\delta y^2_a)$ terms remain on the right hand side of the equations. Thus the equations
are rewritten in the form:
\begin{equation}
\delta y_a-G_{ab} \star ({\cal Y}_b\delta y_b)-\delta f_a=G_{ab} \star\left[ \log(1+ {\cal Y}_b \, \delta y_b)-
{\cal Y}_b \, \delta y_b \right]-\left[ \log(1+\delta y_a)-\delta y_a \right]. \label{iteq1}
\end{equation}
It can be seen that on the left hand side of (\ref{iteq1}) all the quantities are $O(\delta y_a)$, while on the
right hand side all the quantities are $O(\delta y^2_a)$. This separation allows us to define an iterative solution.
If $\delta y_a$s are small then the RHS is a small correction with respect to the LHS, this is why in an iterative 
solution the RHS can be simply taken at the value of the previous iteration.

Let $\delta y^{(n)}_a$ the value of $\delta y_a$ after the $n$th iteration, then $\delta y^{(n+1)}_a$
can be determined from $\delta y^{(n)}_a$ by solving a set of linear integral equations:
\begin{equation}
\delta y^{(n+1)}_a-G_{ab} \star ({\cal Y}_b\delta y^{(n+1)}_b)-\delta f_a=G_{ab} \star\left[ \log(1+ {\cal Y}_b \, \delta y^{(n)}_b)-
{\cal Y}_b \, \delta y^{(n)}_b \right]-\left[ \log(1+\delta y^{(n)}_a)-\delta y^{(n)}_a \right]. \label{iteq2}
\end{equation}
Thus at each step of this iterative method a set of linear integral equations must be solved. Using the discretization
method of the previous subsection, the problem reduces to solving a set of linear algebraic equations, which is a 
straightforward task in numerical mathematics. The very first ($0$th) iteration starts from the asymptotic solution
$\delta y_a=0$ and it corresponds to the solution of the linearized equations, which in our case gives the 
L\"uscher-formula for the energy. 

This (first) method in a certain range of the coupling constant
 defined a numerically convergent iteration to solve the
equations for the ground state of our D-brane anti-D-brane problem, but beyond a certain value of $g$ the method
failed to converge anymore. This is why we worked out a second method, which proved to be much more efficient
than the first one. This efficiency is manifested in two facts. First it converges much faster than the previous iterative 
method, second it gives convergent solutions to our equations even when the solution cannot be accepted as physical 
one\footnote{Beyond a certain value  of the coupling constant the equations in the form presented in section 2.
 are not the right ones anymore, they should be corrected by some new source terms and quantization conditions,
but even for the "wrong" equations the second method shows numerical convergence, giving unacceptable result.}.

This second method can be described simply in words. Instead of defining an iteration as above, we simply take
the discretized version of (\ref{dmodelleq}). We consider it as a set of nonlinear algebraic equations.
As a first step we solve the
linearized discrete equations (i.e. (\ref{iteq2}) with $\mbox{RHS}=0$) and starting from the solution of the linearized equations 
we solve the discrete nonlinear system by Newton-method\footnote{In MATHEMATICA language it can be implemented
by FindRoot[...,Method$\rightarrow$"Newton"].}.

\section{Numerical results}

In this section we summarize our numerical results. We solved numerically the equations
for several integer values of the length parameter $L$. In this section we concentrate
on the states with $L \geq 2$. The $L=1$ special case is discussed in the next section.
For the explanation of the numerical data we will mostly use the $L=2$ case as an example,
because the critical point of this state is the closest one to zero, so it is enough to
work with relatively small values of the coupling constant. This is important
from the numerical point of view, since by increasing $g$ the numerical method becomes
more and more time consuming.

\begin{figure}[htb]
\begin{flushleft}
\hskip 15mm
\leavevmode
\epsfxsize=120mm
\epsfbox{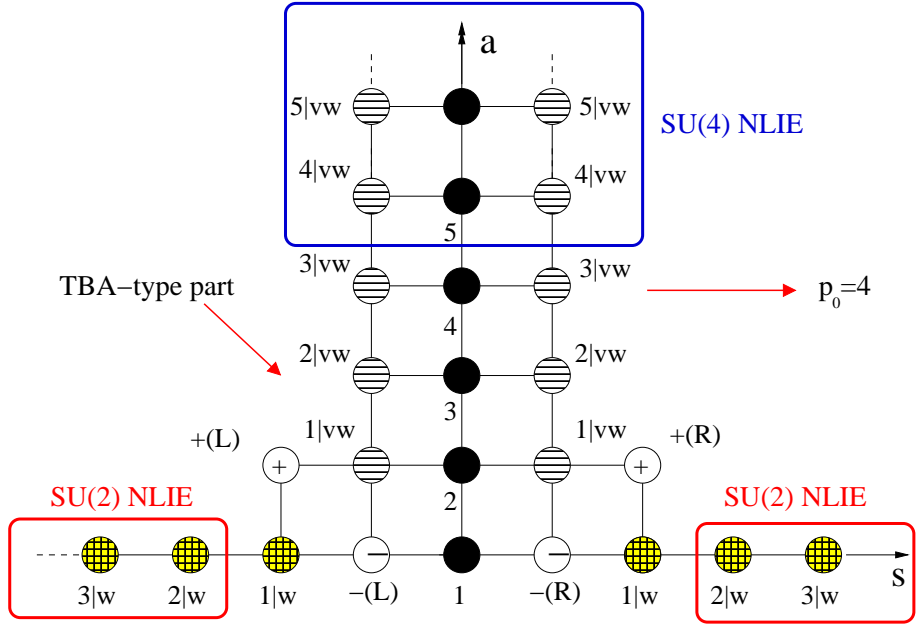}
\end{flushleft}
\caption{{\footnotesize
The pictorial representation of the $Y$-system and the HNLIE structure with the choice $p_0=4$. 
}}
\label{YHNLIE}
\end{figure}

First the parameters of the numerical method is discussed. There are three parameters in the
nonlinear integral equations (\ref{TBAmvw})-(\ref{lasteq}). The most important one is the coupling constant
$g$, then there are two other parameters which allow us to formulate the equations according to our purposes.
 The two parameters are $p_0$ and $C$, where $p_0$ is a kind of "truncation index",
which tells us the node number starting from which the upper TBA equations are replaced by
$SU(4)$ NLIE variables (see figure \ref{YHNLIE}.). The parameter $C$ is a free parameter in the asymptotic solution
for the upper $SU(4)$ NLIE variables (\ref{bfirst}-\ref{dvect}) and it enters the equations such that the asymptotic
solution around which the equations are formulated 
contain this parameter.
From this discussion it is obvious that $g$ is a physical parameter which means that the energy depends on it,
while the other two parameters $p_0$ and $C$ correspond to different formulations of the same mathematical problem,
so the energy does not depend on them. Thus the choice of these parameters is in our hand and we tried to
choose such values for them which allows us numerical convergence in the widest range in $g$.
For example the $C=0$ choice is the best for numerical purposes since due to (\ref{bmMd}) a $u \rightarrow -u$
symmetry arises in the $SU(4)$ HNLIE variables minimizing the number of unknowns in the problem.
Tuning $p_0$ might have two advantages. First, numerical experience shows that for large $p_0$ the Chebyshev
coefficients of the unknowns entering the formula (\ref{OmegaKQ}) for $\Omega$,
decay faster, which allows for higher numerical precision. Second also from
numerics we learn that with $p_0$ fixed at certain values of $g$ non physical results are obtained
from the numerical solution of the problem. This is a consequence of new local singularities entering 
the problem, but we still did not take them into account in the equations. 
These new singularities show up mostly in the $SU(4)$ NLIE variables, thus by increasing the value of
$p_0$ the appearance of such singularities in the equations can be postponed to higher values of $g$.
 
We solved numerically our equations for different values of $L$ and with various values of $p_0$ and $C$, and
in case the numerical result was physically acceptable for all $p_0$ and $C$ we tried, it was also independent of
these parameters within the numerical errors of the method.

So far we discussed the parameters of the continuous integral equations and their role in the numerical solution.
Now we turn to discuss the numerical parameters of the equations. The numerical parameters are artifacts
of the numerical method, and they arise mostly from the discretization method described in section 3.
We note that there is no cutoff parameter in our numerical method, neither in the integration range nor
in the index of $Y$-functions. Everything is treated in an exact manner, the only source of numerical 
errors is the discretization of the unknowns and the convolutions.
Here we give the most used subinterval vectors of our numerical computations. On each subinterval we used
an $l_k=10$ order Chebyshev approximation. The subinterval vector $A_{\pm}$ of $[0,2g]$ is given by the empirical
formula:
\begin{equation}
A_{\pm}=\left\{ \begin{array}{llr}
A_{\pm,<} & \mbox{if } & g\leq 2,\\
A_{\pm,>} & \mbox{if } & g\geq 2,
\end{array}\right.
\end{equation}
where the vectors in components take the form:
\begin{equation}
A_{\pm,<}^{(k)}=\frac{2 g \, k }{[2 g]+1}, \qquad k=1,...,[2g]+1,
\end{equation}
\begin{equation}
A_{\pm,>}^{(k)}=\left\{\frac12,1,\underline{v},2g-\frac34,2g-\frac12,2g-\frac14,2g \right\},
\end{equation}
with $\underline{v}$ having vector components:
\begin{equation}
v_j=1+j \, \frac{2 g-2 }{\left[2 g-\frac32\right]}, \qquad j=1,...,\left[2 g-\frac32\right].
\end{equation}
Here $[...]$ stands for integer part.
The set of subinterval vectors $A_Q$ of $[0,B(Q)]$ could also be given by an appropriate empirical formula,
but it would take such a complicated form, that it is better to write down the requirements from which it
can be constructed\footnote{These requirements are based on numerical experiences with the choice $l_k\geq 10$.}.
The requirements can be formulated in the language of the variable $t \in [0,B(Q)]$. The elements of the vector
 $A_Q$ divide the interval $[0,B(Q)]$ into subintervals. Our requirements constrain the allowed length
of the subintervals with respect their location within the whole interval $[0,B(Q)]$. 
The requirements are as follows:
\begin{itemize}
\item The first element of $A_Q$ is $\frac12$.
\item The length of subintervals $\Delta t$ in the range $\frac12<t<2$ is approximately $\frac13$: $\Delta t\lessapprox \frac13$.
\item The length of subintervals in the range $2<t<3$ is approximately $\frac12$: $\Delta t\lessapprox \frac12$.
\item The length of subintervals in the range $3<t<B(Q)$ is approximately $1$: $\Delta t\lessapprox 1$.
\end{itemize}
In practice the length of the subintervals are slightly "squeezed" with respect to the conditions above
to fill the full $[0,B(Q)]$ properly\footnote{Not to have very small subintervals: $\Delta t\lessapprox 0.1$.}.

Finally, we note that for checking the numerical precision, we also did numerical computations with $l_k=12,14,16$
 keeping the subintervals fixed and also with keeping $l_k=10$, but doubling the number of subinterval points.  

Before turning to present the numerical results we would like to say a few words about the possible tests of the
numerical results. Namely, how one can recognize a wrong result. This is also a very important point of the
numerical method, since there are a lot of equations with very complicated kernels and it is easy to make mistakes
during writing the code of the numerical solution.
There are three basic things that we can check from the numerical results.

The first check is dictated by the energy equation (\ref{eps}). It is known that the energy
starts at the first wrapping order (i.e. $e^{-L}$) and this first order correction is exactly 
given by the L\"uscher formula \cite{DH}:
\begin{equation}
\Delta E(L)=-\sum\limits_{Q=1}^{\infty} \int\limits_{0}^{\infty}  \frac{du}{2 \pi} \,
\frac{d\tilde{p}_Q}{du} \, Y^o_{Q}(u),
\label{Luscher0}
\end{equation} 
with $Y^o_{Q}(u)$ given explicitly in (\ref{YQo}).
This quantity can be computed numerically with any digits of precision, so its value is known exactly at
any values of $g$ and $L$. The L\"uscher-formula (\ref{Luscher0}) corresponds to the linearized
version of our equations (\ref{TBAmvw})-(\ref{lasteq}), this is why solving the linearized set of equations 
(which is the first step for the iterative solution) we should reproduce the numerical evaluation
of (\ref{Luscher0}). This is a nontrivial check on the kernels, on the discretization method and on the equations
themselves as well. In addition since this test is quantitative it can tell some information also
on the numerical precision of the method\footnote{If one experiences that the numerical solution of the 
linearized problem agrees with  the numerical value of (\ref{Luscher0}) within certain digits of precision,
than the deviation from the L\"uscher result can be a good starting estimate to the numerical error.
One cannot expect better accuracy, but the precision will not become much worse either.}.

This test can signal problems on solving the linearized problem. The remaining two tests can signal some discrepancies
during the solution of the nonlinear problem.

The second testing condition is that from the numerical solution $Y_{p_0-2|vw}$ must be real.
This sound trivial, but it is not trivial at all. If one takes a look at the equation (\ref{Yvwp0m2})
of $Y_{p_0-2|vw}$, one can recognize that there are complex quantities on the RHS which do not
form conjugate pairs. So, the reality of the LHS is not guaranteed by the form of the equations,
but it is guaranteed by the form of the solution.
Thus the second testing condition is expressed by the inequality:
\begin{equation}
|\mbox{Im}\log Y_{p_0-2|vw}|\leq \, \mbox{Numerical error}, 
\qquad 10^{-6} \lessapprox \mbox{Numerical error} \lessapprox 10^{-9}.
\end{equation}
Here we wrote the typical numerical errors we had during the computations.

The third test is based on the approximation scheme we use. One must check whether
the Chebyshev coefficients of the unknowns decay as it is expected. From such a check
the numerical precision of the method can be read off and it can shed light on some
anomalous divergent behavior of the numerical solution. Thus it can indicate possible 
errors in the elimination of the divergent $\ln u$ terms in (\ref{separ}) and (\ref{b2bar},\ref{d2bar}).

\begin{figure}[htbp]
\begin{center}
\begin{picture}(260,70)  \epsfxsize=70mm
\put(0,15) {\epsfbox{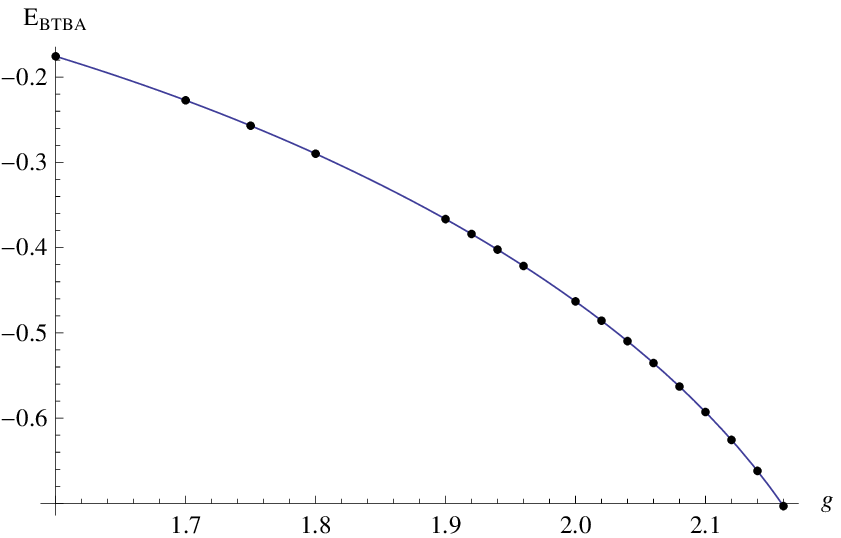}}
\put(100,15) {\epsfbox{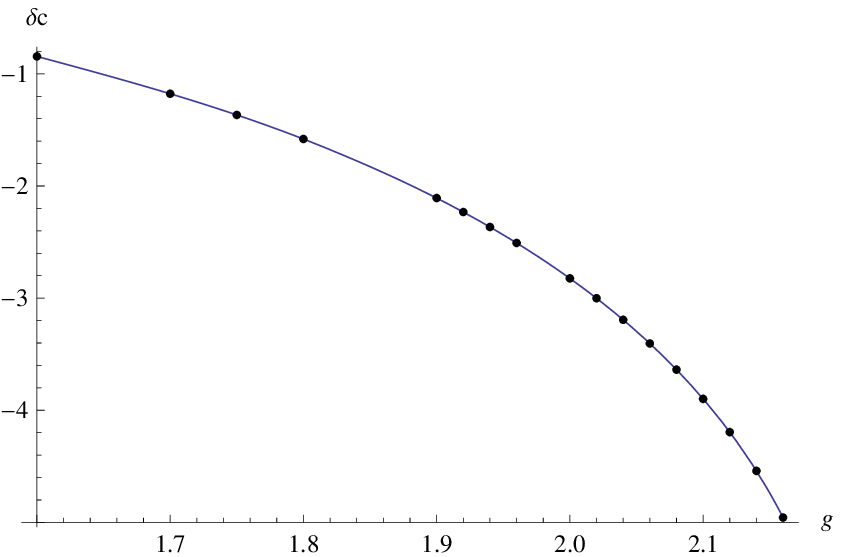}}
\put(0,0) {\parbox{150mm}
{\caption{ \label{aL2}\protect {\small
 $E_{BTBA}$ (on the left) and $\delta c$ (on the right) as functions of $g$ for the $L=2$ state. }}}}
\end{picture}
\end{center}
\end{figure}

The numerical results for the $L=2$ case can be seen in figure \ref{aL2} and table \ref{tL2}. 
In the table we show not only the
energy $E_{BTBA}$ at different values of the coupling $g$, but the constant $\delta c$, as well.
The other columns of the table are related to the solution of the linearized equations;
$E_{BTBA}^{(0)}$ and $\delta c^{(0)}$ are the energy and the global constant from the
numerical solution of the linearized equations. $\Delta E_{BTBA}^{(0)}$ stands for the
deviation of $E_{BTBA}^{(0)}$ from the exact L\"uscher result. This quantity gives some
information on the numerical accuracy of the method. Finally the column "number of nodes"
tells us the cutoff index of the L\"uscher formula, which is necessary to get the L\"uscher energy
with the precision given by $\Delta E_{BTBA}^{(0)}$. This number is not equal to $p_0$ in our equations.
For the $L=2$ state, in case of $0<g<1.9$ we used $p_0=4$, for $1.9<g<2.1$ we used $p_0=8$, and
in the range $2.1<g<2.14$ we took $p_0=12$. Finally at $g=2.16$ we used $p_0=26$ to get acceptable
numerical results. Then beyond this point we could not save our equations
from the entrance of new singularities by increasing the value of $p_0$ with a reasonable $O(10)$ amount.
Because of this reason we could not get really close to the supposed critical point. There $E_{BTBA}\sim -1$,
but we could reach only $E_{BTBA}\sim -0.7$ at $g=2.16$. Apart from this very embarrassing fact, some important
features can be read off from the numerical data. First of all it can be seen that in the range $g<2.16$
the energy is very slowly varying function of $g$, so there is no sign of any divergent behavior.
What is more interesting is the behavior of the global constant $\delta c$. It is negative and it
decreases faster and faster as $g$ is increased. From the definition of $\delta c$ (\ref{separ}) it follows that
all $Y_Q$-functions are proportional to its exponent: $Y_Q \sim \xi=e^{\delta c}$. The  fast decrease of $\delta c$
indicates that though $Y_Q$ has worse and worse large $u$ asymptotic by the increase of $g$, its
global magnitude is actually decreasing. This remark can be understood from the TBA formulation of the
energy.
\begin{equation}
E_{BTBA}=-\sum\limits_{Q=1}^{\infty} \int\limits_{0}^{\infty}  \frac{du}{2 \pi} \, \frac{d \tilde{p}_Q}{du}\,
\log(1+Y_{Q}(u)).
\label{EBTBA}
\end{equation}
Close to the critical point $E_{BTBA}$ is supposed to be finite \cite{DH} $E_{BTBA} \sim 1-L$, but naively
the sum in the RHS of (\ref{EBTBA}) would diverge due to the large $Q$ terms. Since $Y_Q$
is small for large $Q$, in leading order\footnote{For large $Q$.} the $\log(1+Y_Q) \rightarrow Y_Q$ replacement can be done:
\begin{equation}
E_{BTBA}=\underbrace{-\sum\limits_{Q=1}^{Q_0} \int\limits_{0}^{\infty}  \frac{du}{2 \pi} \, \frac{d \tilde{p}_Q}{du}\,
\log(1+Y_{Q}(u))}_\text{Finite}-\, \xi \underbrace{\sum\limits_{Q=Q_0}^{\infty} \int\limits_{0}^{\infty}  \frac{du}{2 
\pi} \, \frac{d \tilde{p}_Q}{du}\, \tilde{Y}_{Q}(u)}_\text{Diverges close to the critical point}+...,
\label{EBTBA1}
\end{equation}
where $Q_0$ is an arbitrary index cutoff scale 
and $Y_Q=\xi \,\tilde{Y}_{Q}$ replacement was applied. Since $\xi$ is $Q$-independent all the
dangerous $Q$ dependence is still in $\tilde{Y}_{Q}$. In (\ref{EBTBA1}) approaching to the critical
point the second sum starts to diverge, and the global multiplicative factor $\xi$ must tend to zero
in order to ensure the finiteness of both sides of the equation. Our numerical data seems to support
this picture. Namely $\delta c \to -\infty$ as going closer and closer to the critical point.

\begin{table}
\begin{center}
\begin{tabular}{|c|c|c||c|c||c|c|}
\hline
$g$ & $E_{BTBA}$ & $\delta c$ & $E_{BTBA}^{(0)}$ & $\delta c^{(0)}$ & $\Delta E_{BTBA}^{(0)}$ & number of nodes\\
\hline
1.6 & -0.175553 & -0.844383 & -0.185898 & -0.893355 & $3.7 \cdot 10^{-6}$ & 25 \\
\hline
1.7 & -0.2271599 & -1.17751 & -0.24183601 & -1.21357 & $4.9 \cdot 10^{-6}$ & 25 \\
\hline
1.75 & -0.25693719 & -1.36622 & -0.2738668 & -1.40256 & $5.7 \cdot 10^{-6}$ & 27 \\
\hline
1.80 & -0.2897776 & -1.58077 & -0.3088130 & -1.61278 & $6.3 \cdot 10^{-6}$ & 28 \\
\hline
1.90 & -0.366494169 & -2.10766 & -0.38810198 & -2.10321 & $7.9 \cdot 10^{-6}$ & 30 \\
\hline
1.92 & -0.38393979 & -2.23237 & -0.40555472 & -2.21339 & $8.2 \cdot 10^{-6}$ & 30 \\
\hline
1.94 & -0.402255118 & -2.36573 & -0.4235649 & -2.32785 & $8.7 \cdot 10^{-6}$ & 30 \\
\hline
1.96 & -0.42147149 & -2.50781 & -0.4421440 & -2.44671 & $9.0 \cdot 10^{-6}$ & 30 \\
\hline
2.00 & -0.46303978 & -2.82377 & -0.4810544 & -2.69809 & $9.9 \cdot 10^{-6}$ & 31 \\
\hline
2.02 & -0.48564199 & -3.00085 & -0.5014098 & -2.83086 & $9.9 \cdot 10^{-6}$ & 32 \\
\hline
2.04 & -0.50966430 & -3.19333 & -0.52237993 & -2.96847 & $1.0 \cdot 10^{-5}$ & 32 \\
\hline
2.06 & -0.53532776 & -3.40422 & -0.5439774 & -3.11107 & $1.0 \cdot 10^{-5}$ & 32 \\
\hline
2.08 & -0.56291307 & -3.63744 & -0.566214 & -3.25878 & $1.0 \cdot 10^{-5}$ & 33 \\
\hline
2.10 & -0.592805 & -3.89861 & -0.589106 & -3.41179  & $8.9 \cdot 10^{-6}$ & 35 \\
\hline
2.12 & -0.625515 & -4.19506 & -0.612655 & -3.56999 & $1.2 \cdot 10^{-5}$ & 34 \\
\hline
2.14 & -0.661868 & -4.54055 & -0.636888 & -3.73339 & $9.6 \cdot 10^{-6}$ & 36 \\
\hline
2.16 & -0.7031687 & -4.956683 & -0.661809 & -3.90338 & $7.0 \cdot 10^{-6}$ & 39 \\
\hline
\end{tabular}
\bigskip
\caption{Numerical data for the $L=2$ state.}
\label{tL2}
\end{center}
\end{table}
\normalsize

\begin{figure}[htb]
\begin{flushleft}
\hskip 15mm
\leavevmode
\epsfxsize=120mm
\epsfbox{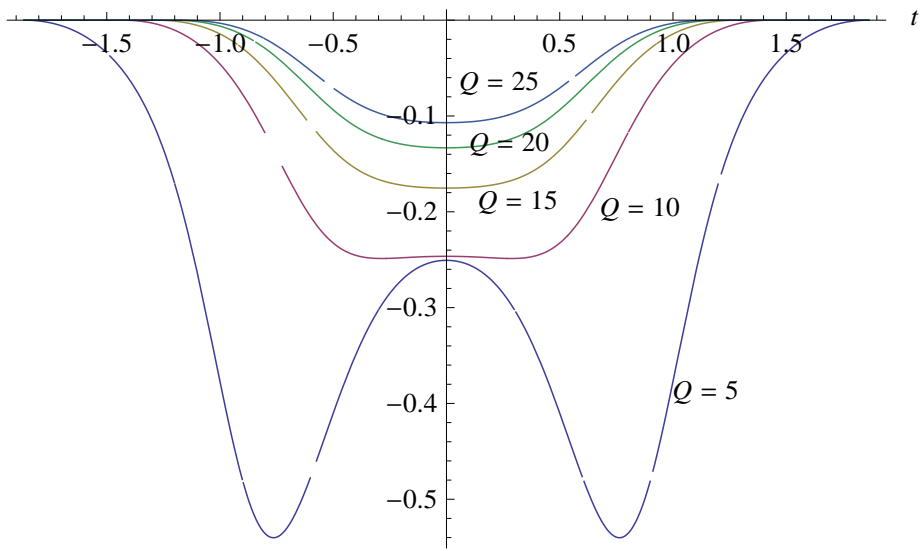}
\end{flushleft}
\caption{{\footnotesize
Large $Q$ behavior of $\delta {\cal F}_Q$ from numerical data at $g=2.16$ with $p_0=26$. 
}}
\label{1}
\end{figure}

 In \cite{DH} from $Y$-system
arguments the large $Q$ behavior of $Y_Q$ was also estimated by the formula:
\begin{equation}
Y_Q(u)\simeq \xi(g) \, \frac{1}{\left( u^2+\frac{Q^2}{g^2}\right)^{2 \, E_{BTBA}}} \, Y^o_Q(u), 
\qquad \xi=e^{\delta c},
\label{largeYQ}
\end{equation}  
where $\delta c$ is defined after (\ref{separ}) in section 2. (\ref{largeYQ}) is a very important formula, because
it plays crucial role in the analytical determination of the critical point. Since the numerical solution of the
 HNLIE equations of section 2. does not require the introduction of any index cutoff, it takes into account the
contributions of all the $Y$-functions of the infinite $Y$-system. This makes it possible to test numerically the
correctness of the large $Q$ estimate (\ref{largeYQ}). In case (\ref{largeYQ}) holds, it implies that 
$\delta \ln \bar{y}_Q=\ln \bar{y}_Q-\ln \bar{y}_Q^{o}$ tends to zero as $1/Q$ for large $Q$. In figure \ref{1}. the numerical
demonstration of this statement can be seen. The plotted functions are defined by the formula:
\begin{equation}
\delta{\cal F}_{Q}(t)=\left\{ \begin{array}{llr}
\delta \ln \bar{y}_Q\left(x_Q(B(Q)-t)\right) & \mbox{if } & t> 0,\\
\delta \ln \bar{y}_Q\left(-x_Q(B(Q)+t)\right) & \mbox{if } & t< 0,
\end{array}\right. \label{dFQ}
\end{equation}
where $x_Q(t)=Q \left( \frac{B(Q)}{t}-1\right), \quad B(Q)=\frac{2g}{Q}+1$.
The plots of figure \ref{1}. are based on the numerical computation with $p_0=26$ at $g=2.16$.
Figure \ref{1}. nicely demonstrates the expected $1/Q$ behavior of the functions $\delta{\cal F}_Q$.

\begin{figure}[htb]
\begin{flushleft}
\hskip 15mm
\leavevmode
\epsfxsize=120mm
\epsfbox{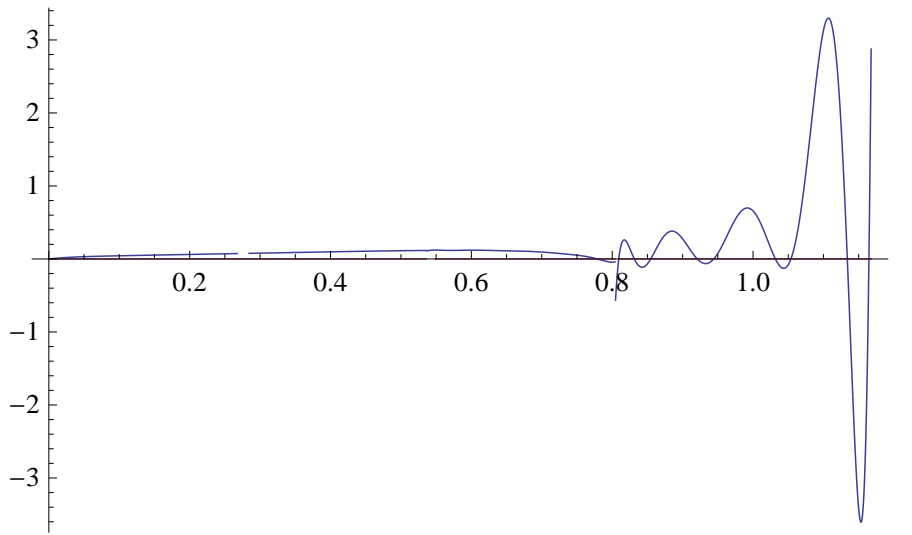}
\end{flushleft}
\caption{{\footnotesize
The anomalous behavior of $\mbox{Im}\log(1+\delta b_6)$ at $g=2.18$ and $p_0=26$. 
}}
\label{2}
\end{figure}

For the $L=2$ state beyond $g=2.16$ the numerical solution of the discretized problem
did not give physically acceptable results. To get some insight into the source of the
problems, at $g=2.18$ we plotted the imaginary part of the LHS of the last equation in (\ref{bA}), namely
$\mbox{Im}\log(1+\delta b_6)$ at $u=x_{p_0}(B(p_0)-t)$. Figure \ref{2}. shows that there is 
a jump of $2 \pi$, when $t$ is close to $B(p_0)$. (I.e. large u.)\footnote{Here the sampling points are
connected according to the Chebyshev approximation. This is why the jump of the logarithm is not "sharp".}
 This fact shows us that
the equations we solved numerically are not the right ones anymore. Something is missing from the
equations. Either a special object \cite{Destri:1997yz,Hegedus:2007jw} or some other local singularities of the $T$- and 
$Q$-functions of the problem, which enter those strips of the complex plane, which are relevant
in the derivation of the HNLIE equations. 

The numerical data for the $L=3$ and $L=4$ states are given by table \ref{L3} and \ref{L4}. Also in case of these
states the appearance of new singularities obstacled us to get close to the critical point in the
framework of the HNLIE technique.

\begin{table}
\begin{center}
\begin{tabular}{|c|c|c||c|c||c|}
\hline
$g$ & $E_{BTBA}$ & $\delta c$ & $E_{BTBA}^{(0)}$ & $\delta c^{(0)}$ & $\Delta E_{BTBA}^{(0)}$ \\
\hline
2.2 & -0.114591 & -0.62945 & -0.12907 & -0.711869 & $8.4 \cdot 10^{-7}$  \\
\hline
2.6 & -0.21909 & -1.33443 & -0.267823 & -1.641000 & $3.0 \cdot 10^{-7}$  \\
\hline
2.8 & -0.286833 & -1.82583 & -0.366547 & -2.34855 & $2.1 \cdot 10^{-6}$  \\
\hline
3.0 & -0.365866 & -2.42503 & -0.488968 & -3.26271 & $2.9 \cdot 10^{-6}$  \\
\hline
3.2 & -0.457294 & -3.14677 & -0.638504 & -4.42131 & $8.7 \cdot 10^{-6}$  \\
\hline
3.4 & -0.56282 & -4.01232 & -0.818842 & -5.86636 & $1.1 \cdot 10^{-5}$  \\
\hline
3.6 & -0.685108 & -5.05271 & -1.03391 & -7.64329 & $1.5 \cdot 10^{-5}$  \\
\hline
\end{tabular}
\bigskip
\caption{Numerical data for the $L=3$ state.}
\label{L3}
\end{center}
\end{table}
\normalsize

\begin{table}
\begin{center}
\begin{tabular}{|c|c|c||c|c||c|}
\hline
$g$ & $E_{BTBA}$ & $\delta c$ & $E_{BTBA}^{(0)}$ & $\delta c^{(0)}$ & $\Delta E_{BTBA}^{(0)}$ \\
\hline
2.6 & -0.0716174 & -0.427755 & -0.0793412 & -0.476413 & $6.0 \cdot 10^{-7}$  \\
\hline
2.8 & -0.0975242 & -0.607523 & -0.111564 & -0.699788 & $1.1 \cdot 10^{-8}$  \\
\hline
3.0 & -0.128116 & -0.829046 & -0.151888 & -0.991352 & $4.2 \cdot 10^{-8}$  \\
\hline
3.2 & -0.163439 & -1.0949 & -0.201422 & -1.36341 & $9.2 \cdot 10^{-8}$  \\
\hline
3.4 & -0.203514 & -1.40733 & -0.261362 & -1.82949 & $9.8 \cdot 10^{-8}$  \\
\hline
3.6 & -0.24835 & -1.76832 & -0.332987 & -2.40437 & $7.0 \cdot 10^{-7}$  \\
\hline
4.0 & -0.35239 & -2.64384 & -0.51686 & -3.94631 & $8.6 \cdot 10^{-7}$  \\
\hline
4.2 & -0.411691 & -3.16247 & -0.632106 & -4.9496 & $2.7 \cdot 10^{-6}$  \\
\hline
4.4 & -0.545354 & -4.3733 & -0.917326 & -7.52081 & $1.3 \cdot 10^{-6}$  \\
\hline
\end{tabular}
\bigskip
\caption{Numerical data for the $L=4$ state.}
\label{L4}
\end{center}
\end{table}
\normalsize

\section{Comments on the $L=1$ case}

The $L=1$ ground state is mysterious, since so far the anomalous dimension of this state 
could not be determined even for small $g$ either from field theory or from integrability considerations
\cite{DH}.
Here we concentrate on the integrability side. There the boundary L\"uscher formula \cite{Correa:2009mz,Bajnok:2010ui}
 diverges for this state \cite{DH}.
For generic $L$ the L\"uscher formula is simply the expansion of the TBA energy formula
around the asymptotic solution with the replacement: $\log(1+Y_Q) \rightarrow Y_Q^o$. For
small coupling it takes the form \cite{DH}:
\begin{equation}
\Delta E(L)=-\sum\limits_{Q=1}^{\infty} \int\limits_{0}^{\infty}  \frac{du}{2 \pi} \,
\frac{d\tilde{p}_Q}{du}\,
Y^o_{Q}\left(u\right)\simeq -\left(\frac{g}{2}\right)^{4L} \, \left\{   \frac{4}{4L-1} \,{4L \choose 2L} \, \zeta(4L-3)+O(g^2)\right\}.
\label{Luscher}
\end{equation}
This small coupling expression diverges for $L=1$, since this point sits exactly on the pole of the $\zeta$
-function. As for the origin of this divergence; in (\ref{Luscher}) the individual integrals are convergent,
 but their sum for $Q$ causes the divergence. 
In \cite{DH} it was argued that also for any larger $L$ the TBA energy formula would diverge
beyond a certain critical value of the coupling: $g_c(L)$. Assuming that the energy is a monotonously
decreasing function of $g$, which is supported by numerical results, this critical point can be expressed
clearly in terms of the energy by the criterion:
\begin{equation}
E_c(L)\equiv E(g_c(L))=1-L.
\end{equation}
In \cite{DH} this point was interpreted as a turning point where the energy becomes imaginary and as a 
physical consequence the ground state becomes tachyonic. For the $L=1$ state the critical point is right
at $g=0$ assuming that for small $g$ the energy is also small.

Now let us turn our attention to the HNLIE description of the problem detailed in section 2. 
Here there are no infinite sums and even for $L=1$ all the convolutions of the integral equations
seem to converge\footnote{If we assume that large $u$ behavior of the unknown
functions, which was used to derive the BTBA equations from discontinuity relations and Y-system.}.
For the first sight there is no sign of any problem in the HNLIE description and it seems that only the
TBA description is inappropriate to treat the $L=1$ case.
But unfortunately this is not the case.

 We can write down the discretized integral equations
for the $L=1$ case as well, and using the Newton-method, we can solve them for small values of the
coupling\footnote{Typically $g \sim 10^{-1}$.}. We always get some numerical solution for the
discretized problem, but it turns out that the Chebyshev coefficients of the unknowns, which correspond
to the large $u$ subinterval do not form a decaying series. This phenomenon is a typical sign of some
weak (probably logarithmic) large $u$ divergence of the unknowns. If one increases the number of subintervals
and sampling points the situation remains the same. The conclusion is that we can solve the discretized
problem, but the solution cannot be interpreted as the discretely approximated version of the 
continuous solution of our integral equations. In other words the continuous HNLIE equations have no
solution for $L=1$.

In order to get some analytical insight why the solutions become diverging at large $u$ 
let us consider the TBA formulation of the problem ($p_0 \to \infty$ in HNLIE).
It is known \cite{DH} that the TBA energy comes from the coefficient of the most divergent
$\log|u|$ term in the large $u$ expansion of $\log Y_Q$:
\begin{equation}
\log Y_Q(u) = -4(L+E_{BTBA}) \, \log|u|+O(1).
\end{equation}
The $E_{BTBA}$ term originates from the RHS of the TBA equations for $\log Y_Q$ from the
convolution term $\sum\limits_{Q'=1}^{\infty} \log(1+Y_{Q'}) \star K^{Q'Q}_{{\mathfrak{sl}(2)}}$ by exploiting
the large $u$ expansion of the kernel: 
$K^{Q'Q}_{{\mathfrak{sl}(2)}}(v,u)=-\frac{1}{\pi} \, \frac{d\tilde{p}_{Q'}}{dv}\,
\log|u|+O(1)$. $K^{Q'Q}_{{\mathfrak{sl}(2)}}$ has better large $Q'$ behavior than that of $\frac{d\tilde{p}_{Q'}}{dv}$,
since it behaves like $1/Q'$. As a consequence contrary to the energy formula,
 the sum of dressing convolutions is convergent indeed.
Thus one might think that for $L=1$ the problem emerges, because 
for the derivation of the energy formula we expanded the sum of dressing convolutions
term by term for large $u$. This is why instead of this usual procedure,
we consider the sum of dressing convolutions itself,
compute it and then at the end of the computation we take the large $u$ expansion.
This procedure is carried out in the small coupling limit.
We need the leading order small coupling expression of the dressing kernel 
in the mirror-mirror channel\footnote{Here we use the
rapidity convention where the branch points are at $\pm 2 g$.}:
\begin{equation}
\begin{split}
K^{Q'Q,\,{(0)}}_{{\mathfrak{sl}(2)}}(u_1,u_2)=&-\frac{1}{2 \pi} \left[ 
\psi\left( 1+\frac{Q'}{2}-i \frac{i}{2} u_1 \right)+
\psi\left( 1+\frac{Q'}{2}+i \frac{i}{2} u_1 \right) \right. \\
&\left. -\psi\left( 1+\frac{Q'+Q}{2}+i \frac{i}{2} (u_2-u_1) \right)-
\psi\left( 1+\frac{Q'+Q}{2}-i \frac{i}{2} (u_2-u_1) \right)
\right]+....
\end{split}
\end{equation}
Then the formula, the large $u$ expansion of which accounts for the small coupling expanded energy,
is given by:
\begin{equation}
{\cal O}(u,Q)=\sum\limits_{Q'=1}^{\infty} \, \int\limits_{-\infty}^{\infty} \, \frac{dv}{4 \pi} \,
Y^{o,{(L=1)}}_{Q'}(v) \, K^{Q'Q,\,{(0)}}_{{\mathfrak{sl}(2)}}(v,u), 
\end{equation}
where $Y^{o,{(L=1)}}_{Q}(u)=g^4 \, \frac{16\,Q^2 \, u^2}{(u^2+Q^2)^{3}}$ is the leading small coupling expression
of (\ref{YQo}) at $L=1$. The second derivative of ${\cal O}(u,Q)$ can be computed explicitly by simple Fourier
space technique. We take the Fourier form of each functions under integration, the convolution is the product
of the individual Fourier transforms, the sum for $Q'$ can be easily done in Fourier space and at the end of the
process everything is transformed back to the $u$ space. In such a manner one gets a bulky, but explicit expression
for $\frac{d^2}{du^2}{\cal O}(u,Q)$, which we do not present here, only its large $u$ expansion:
\begin{equation}
\frac{d^2}{du^2}{\cal O}(u,Q)=4 \, g^4 \, \left(\frac32-\gamma_E+2 \ln 2 \right) \, \frac{1}{u^2}-8 \, g^4\,
 \frac{\log u}{u^2}+O(\frac{1}{u^3}).
\end{equation}
Integrating twice the large $u$ expansion at small coupling becomes:
\begin{equation}
{\cal O}(u,Q)=4 \, g^4 \, \left(\frac12+2 \, \gamma_E- \ln 4 \right) \log u+ 4 \, g^4 \, (\log u)^2+... \label{O}
\end{equation}
From (\ref{O}) it is obvious why the naive L\"uscher energy formula diverged. Because the leading order
large $u$ term is not the expected $\sim \log |u|$, but $\sim (\log u)^2$.
This is the key point of the problem, since in this case after this first iteration
$Y_Q$ acquires an unwanted type of large $u$ term, which makes $Y_Q$ divergent for large $u$:
\begin{equation}
Y_Q(u)\sim u^{\left(-4L+4 \, g^4 \, \left(\frac12+2 \, \gamma_E- \ln 4 \right)+...\right)} \, e^{4 \,g^4 \, (\log u)^2+...}.
\end{equation}
This large $u$ divergence contradicts to what was assumed about the large
$u$ behavior of $Y_Q$ at the derivation of the integral equations, since it was supposed to decay. 
In this example we have shown in the small coupling limit, that during the iterative solution of the
BTBA equations, $\log Y_Q$  acquires an extra $\sim (\log |u|)^2$ behavior at infinity, which made
$Y_Q$ an exploding function at infinity. This means that the iterative solution of the TBA equations
leaves the  class of physically acceptable solutions.


One might ask the question, whether it is possible to keep somehow the qualitative 
large $u$ behaviors that we assumed at the derivation of the equations?
Here we sketch a possible idea for small coupling to the $L=1$ case. 

Let us assume that we managed to modify the TBA equations, such that all $Y$-functions have the
large $u$ behavior we want. Since most of the TBA equations reflect the structure of the 
$Y$-system functional equations we expect to modify only those equations which are affected by also the discontinuity 
relations. It follows, that for large $Q$, the formula for the estimate for $Y_Q$ (\ref{largeYQ}) remains the same.
Now, we assume that for small $g$ the energy is also small and take the simultaneous small $g$ and small
energy expansion of the RHS of the TBA energy formula (\ref{EBTBA}). In leading order the large $Q$ terms will dominate:
\begin{equation} 
\begin{split}
E_{BTBA}&\simeq-\sum_{Q=1}^{\infty} \int\limits_{-\infty}^{\infty}\frac{du}{4 \, \pi} \, \hat{Y}_Q\left(\frac{u}{g}\right) =
-\tilde{\xi} \sum_{Q=1}^{\infty}\int\limits_{-\infty}^\infty\frac{du}{4\pi}
\left(g^{2}\right)^{2L}16 Q^{2}\frac{u^{2}}{(u^{2}+Q^{2})^{2(L+E_{BTBA})+1}}  \\
&=-\frac{\tilde{\xi}}{2^{4(L+E_{BTBA})}} \, \frac{4g^{4L}}{4(L+E_{BTBA})-1}{4(L+E_{BTBA}) \choose 2(L+E_{BTBA})} \sum\limits_{Q=1}^{\infty} \frac{1}{Q^{(4(L+E_{BTBA})-3)}} \, \\
&=-\frac{\tilde{\xi}}{2^{4(L+E_{BTBA})}} \, \frac{4g^{4L}}{4(L+E_{BTBA})-1}{4(L+E_{BTBA}) \choose 2(L+E_{BTBA})} \zeta (4(L+E_{BTBA})-3) \, \\
&\simeq -\frac{\tilde{\xi} \, g^4}{8 \,E_{BTBA}}+O(\tilde{\xi} g^4),
\end{split}\label{Espec1}
\end{equation}
where $\hat{Y}_Q$ denotes the large $Q$ estimate (\ref{largeYQ}) of $Y_Q$, $\tilde{\xi}=\xi \, g^{4 E_{BTBA}}$ as a consequence of the
$u \rightarrow u/g$ change of variables and the pole term in $E_{BTBA}$ comes from the pole of the
$\zeta$-function.
In our HNLIE approach the energy $E_{BTBA}$ and the constant $\delta c$ are parts of 
the equations which means that they are not simply expressed by explicit formulas based on the 
solution of the equations, but must me obtained by solving the set of non-trivially entangled equations.
In this sense (\ref{Espec1}) defines an equation for $E_{BTBA}$ for small $g$. Its leading order solution is:
\begin{equation}
E_{BTBA}=g^2 \, \sqrt{-\tilde{\xi}}+\dots. \label{Eim1}
\end{equation}
If $\tilde{\xi}>0$ then $E_{BTBA}$ becomes imaginary as it would be expected from string-theory expectations \cite{DH}.
To decide the sign of $\tilde{\xi}$
, the equation (\ref{chybrid}) has to be analyzed in the context 
of the small $g$ and $E_{BTBA}$ expansion. 
It turns out that $\tilde{\xi}$ is positive and $O(1)$ for small $g$, so according to (\ref{Eim1})
$E_{BTBA}$ is imaginary. Another remarkable fact is that according to (\ref{Eim1}) $E_{BTBA}$ starts
at $O(g^2)$ instead of the $O(g^4)$ prediction of the boundary L\"uscher formula (\ref{Luscher}). 
This might be another explanation why the coefficient of $g^4$ diverges in the L\"uscher formula for the $L=1$
case. Finally, we note that in the small $g$ and $E_{BTBA}$ expansion of the $L=1$ state, the energy is pure
imaginary only at leading order in $g$, but in higher orders it acquires real part as well.

For the first sight, it might seem that without modifying the equations one immediately gets imaginary energy
when going through the critical point. But, the situation is a bit more subtle. There is a hidden tacit 
modification of the equations.
This is realized in (\ref{Espec1}) by the
replacement:
\begin{equation*}
\sum\limits_{Q=1}^{\infty} \, \frac{1}{Q^{4(L+E_{BTBA})-3}} \rightarrow \zeta(4(L+E_{BTBA})-3).
\end{equation*}
For the $L=1$ case it is an identity for $\mbox{Re}(E_{BTBA})>0$, but for $\mbox{Re}(E_{BTBA})<0$ 
it is not an identity anymore, but a nontrivial analytical continuation in $E_{BTBA}$. 

Such an analytical continuation would require the exact determination of complicated sums of 
convolutions of the TBA equations as functions of the energy. Since this does not seem to be feasible
in practice, we give such an alternative modification of the TBA equations which preserves the
infinite sum structure of the equations, but the sums will converge everywhere for $\mbox{Re}(E_{BTBA})> -L$
except at the critical value $E_{cr}=1-L$.

The basic idea of the modification comes from the sum representations of the $\zeta$-function.
The usual one converges for $\mbox{Re}(s)>1$:
\begin{equation}
\label{zeta1}
\zeta(s)=\sum\limits_{Q=1}^{\infty} \, \frac{1}{Q^s}, \qquad \mbox{Re}(s)>1,
\end{equation}
but there is another representation which converges for $\mbox{Re}(s)>0$:
\begin{equation}
\label{zeta2}
\zeta(s)=\frac{1}{s-1}\sum\limits_{Q=1}^{\infty} \, \left( \frac{Q}{(Q+1)^s}-\frac{Q-s}{Q^s} \right), \qquad \mbox{Re}(s)>0.
\end{equation}
Then the original TBA equations are modified through their infinite sums by the replacements:
\begin{equation}
\sum\limits_{Q=1}^{\infty} \, L_Q \star {\cal K}_Q \rightarrow \frac{1}{s_E-1}\sum\limits_{Q=1}^{\infty} \, 
\left\{ Q \cdot (L_{Q+1} \star {\cal K}_{Q+1})-(Q-s_E) \cdot (L_Q \star {\cal K}_Q)\right\},
\end{equation}
where $s_E=4(L+E_{BTBA})-3$. Taking into account the large $Q$ behavior of all $Y_Q$ functions and all the kernels
of the infinite sums of the TBA equations, the new representation will converge for $\mbox{Re}(E_{BTBA})>-L$. 
This slight modification of the TBA equations might make it possible to go beyond the critical point and get solution
of the TBA equations with large $u$ asymptotics being in accordance with the ones used for the derivation of the equations.

The conclusion of this heuristic argument is that to keep the expected\footnote{This primarily means that
$\log Y_Q \sim \log|u|$ for large $u$, while other $Y$-functions tend to constant.}
qualitative large $u$ behavior
a nontrivial modification of the TBA equations must be carried out, which might lead to complex
energies.

\section{Summary and conclusions}

In this paper  we studied the ground state energy of a pair of open strings stretching 
between a coincident $D3$-brane anti-$D3$-brane pair
in $S^5$ of $AdS_5 \times S^5$. The main motivation for the study is that string-theory predicts that 
the ground state of such a configuration becomes tachyonic for large values of the 't Hooft coupling \cite{DH}.

In \cite{DH} it was shown that the usual integrability based BTBA approach always give real energies for the
ground state and it breaks down at latest when the energy gets close to the critical value:
$E_c(L)=1-L$. This point was interpreted in \cite{DH} as a transition point where the ground state becomes
tachyonic.

Approaching this critical point the contribution of all the $Y$-functions of the BTBA becomes quantitatively
relevant, thus the numerical solution of the truncated BTBA equations cannot give accurate results close to the
critical point. To resolve this difficulty and get more accurate numerical results we transformed the
previously proposed BTBA equations into finite component HNLIE equations. The HNLIE equations were solved
at different values of $g$ and $L$ and the numerical results confirmed the earlier BTBA data. 

During the numerical solution of the HNLIE equations the usual iterative methods failed to converge,
this is why we worked out two numerical methods to reach convergence. The most effective one is, if one
transforms the integral equations into discrete nonlinear algebraic equations and solves them by 
Newton-method. The power of this method is demonstrated by the fact that it gives convergent results even if the
numerical solution is not physically acceptable.
 
Unfortunately, in our numerical studies we could not get very close to the critical point, because new
singularities entered the HNLIE equations taking into account of which would have required an enormous amount
of additional work. Nevertheless, in the range where we could get physically acceptable results, the precision of the
HNLIE data were higher than those of BTBA and the HNLIE approach could give a deeper understanding
of the problem.

For the ground state of the $L=1$ state the critical point is right
at $g=0$ and neither perturbative field theory computations nor the boundary L\"uscher formula
could provide a finite quantitative answer to the anomalous dimension.
Even in this special case the numerical solution of the HNLIE
equations was possible. The results showed that without an appropriate modification of the equations, they
cannot give physically acceptable results. In this case, it means that the solution of the dicretized
problem cannot be considered as a discretized solution of the continuous nonlinear integral equations.
Moreover the large rapidity behavior of the numerical solution is incompatible with the one assumed for
the derivation of the equations. This phenomenon is analytically analyzed in the framework of BTBA and
an idea is sketched to preserve the expected large rapidity behavior of the unknowns. This method is based
on an appropriate modification of the TBA equations 
which would lead to complex energies beyond the critical point.

Hopefully the $L=1$ case at $g=0$ could be treated analytically in the framework of the quantum spectral curve
method \cite{Gromov:2013pga,Gromov:2014caa,Volinuj}, solving the mystery of this state in the context of integrability.

\vspace{1cm}
{\tt Acknowledgements}

\noindent 
The author thanks Nadav Drukker, L\'aszl\'o Palla and J\'anos Balog for useful discussions.
This work was supported by the J\'anos Bolyai Research Scholarship of the Hungarian Academy of
Sciences and OTKA K109312. The author also would like to thank the support of an MTA-Lend\"ulet Grant,
and the Hungarian-French  bilateral grant T\'ET-12-FR-1-2013-0024.
Finally, the author appreciates APCTP for its hospitality where part of this work was
done.

\appendix


\section{Notations, kinematical variables, kernels}

Throughout the paper we use the basic notations and TBA kernels of ref. \cite{AFS}, which we summarize
below.
For any function $f$, we denote $f^{\pm}(u)=f(u\pm \frac{i}{g})$ and in general
$f^{[\pm a]}(u)=f(u\pm \frac{i}{g} a)$, where the relation between $g$ and the 't Hooft coupling
$\lambda$ is given by $\lambda=4 \pi^2 g^2$.
Most of the kernels and also the asymptotic solutions of the HNLIE-system are expressed
in terms of the function $x(u)$:
\begin{equation}
x(u)=\frac12 (u-i\sqrt{4-u^2}), \qquad \mbox{Im} \, x(u)<0,
\end{equation} 
which maps the $u$-plane with cuts $[-\infty,-2] \cup [2,\infty]$ onto the physical region
of the mirror theory, and in terms of the function $x_s(u)$
\begin{equation}
x_s(u)=\frac{u}{2} \left(1+\sqrt{1-\frac{4}{u^2}}  \right), \qquad |x_s(u)|\geq 1,
\end{equation}
which maps the $u$-plane with the cut $[-2,2]$ onto the physical region of the string theory.
Both functions satisfy the identity $x(u)+\frac{1}{x(u)}=u$ and they are related by the $x(u)=x_s(u),$ and 
$x(u)=1/x_s(u)$ relations on the lower and upper half planes of the complex plane respectively.




The momentum $\tilde{p}^Q$ and the energy $\tilde{\cal{E}}_Q$ of a
mirror $Q$-particle are expressed in terms of $x(u)$ as follows:
\begin{eqnarray}
 \tilde{p}_Q(u)=g x\big(u-\frac{i}{g}Q\big)-g
x\big(u+\frac{i}{g}Q\big)+i Q\, ,
~~~~~\tilde{\cal{E}}_Q(u)=\log\frac{x\big(u-\frac{i}{g}Q\big)}{x\big(u+\frac{i}{g}Q\big)}\,.
 \end{eqnarray}
Two different types of convolutions appear in the HNLIE
equations. These are:
\begin{eqnarray}
\nonumber &&f \star {\cal K}(v) \equiv \int_{-\infty}^\infty\, du\,
f(u) \, { \cal K}(u,v)\,, \quad f \, {\, \hat{\star} \,}{\cal K}(v)
\equiv \int_{-2}^2\, du\, f(u) \, {\cal K}(u,v)\,. \label{convs}
\end{eqnarray}
The kernels and kernel vectors entering the HNLIE equations can
be grouped into two sets. The kernels from the first group are
functions of only the difference of the rapidities, thus actually 
they depend on a single variable.
The other group of kernels composed of those, which are not of difference type.

We start with listing the kernels depending on a single variable:
\begin{alignat}{2}
s (u) & = \frac{1}{2 \pi i} \, \frac{d}{du} \log \tau^-(u)= {g \over 4\cosh {\pi g u \over 2}}\,,\quad \tau(u)=\tanh[
\frac{\pi g}{4} u ]\,,
\nonumber \\
K_Q (u) &= \frac{1}{2\pi i} \, \frac{d}{du} \, \log S_Q(u) = \frac{1}{\pi} \, \frac{g\, Q}{Q^2 + g^2
u^2}\,,\quad S_Q(u)= \frac{u - \frac{iQ}{g}}{u + \frac{i Q}{g}} \,, \nonumber\\
K_{MN}(u) &= \frac{1}{2\pi i} \, \frac{d}{du} \, \log
S_{MN}(u)=K_{M+N}(u)+K_{N-M}(u)+2\sum_{j=1}^{M-1}K_{N-M+2j}(u)\,,\nonumber\\
S_{MN}(u) &=S_{M+N}(u)S_{N-M}(u)\prod_{j=1}^{M-1}S_{N-M+2j}(u)^2 =S_{NM}(u)\,. \label{sKQ}
\end{alignat}
The fundamental building block of kernels which are not of difference type is:
\begin{eqnarray}
 K(u,v) = \frac{1}{2 \pi i} \, \frac{d}{du} \, \log S (u,v) = \frac{1}{2 \pi i} \,
\frac{ \sqrt{4-v^2}}{\sqrt{4-u^2}}\, {1\over u-v} \,,\ \  S(u,v)=\frac{x(u) - x(v)}{x(u) x(v) - 1}\,.~~~
\label{Kuv}
 \end{eqnarray}
Using the kernels $K(u,v)$ and $K_Q(u-v)$ it is possible to define a series of kernels which are connected
to the fermionic $Y_{\pm}$-functions.
They are:
\begin{eqnarray}
K_{Qy}(u,v)&=&K(u-\frac{i}{g}Q,v)-K(u+\frac{i}{g}Q,v)\,, \label{KQy}\\
K^{Qy}_\mp(u,v)&=&{1\over 2}\Big( K_Q(u-v) \pm  K_{Qy}(u,v)\Big) \label{KQypm}
\end{eqnarray}
and
\begin{eqnarray}
K_{yQ}(u,v)&=&K(u,v+{i\over g}Q)-K(u,v-{i\over g}Q), \label{KyQ}  \\
K^{yQ}_\pm(u,v) &=& {1\over 2}\Big(K_{yQ}(u,v)\mp K_Q(u-v)\Big)\, . \label{KyQpm}
\end{eqnarray}
Further important kernels entering the $Y_{\pm}$ related TBA-type equations
are defined as follows:
\begin{eqnarray}
\nonumber
K_{xv}^{QM}(u,v) &=&{1\over 2\pi i}{d\over du}\log S_{xv}^{QM}(u,v)\,,\\
\nonumber S_{xv}^{QM}(u,v) &=&
\frac{x(u-i{Q \over g })-x(v+i{M \over g})}{x(u+i{Q \over g })-x(v+i{M
\over g})}\, \frac{x(u-i{Q \over g })-x(v-i{M \over g})}{x(u+i{Q \over g
})-x(v-i{M \over g})}\, \frac{x(u+i{Q \over g })}{x(u-i{Q \over g
})}~~~~\\ &\times
&\prod_{j=1}^{M-1}\frac{u-v-\frac{i}{g}(Q-M+2j)}{u-v+\frac{i}{g}(Q-M+2j)}. \label{Sxv}
\\\nonumber
\end{eqnarray}
The kernels entering the right hand sides of the equation (\ref{hybrid}) for $Y_1$ are
\begin{eqnarray}
\nonumber
K_{vwx}^{QM}(u,v) &=&{1\over 2\pi i}{d\over du}\log S_{vwx}^{QM}(u,v)\,,\\
\nonumber S_{vwx}^{QM}(u,v) &=&
\frac{x(u-i{Q \over g })-x(v+i{M \over g})}{x(u-i{Q \over g })-x(v-i{M\over g})}\,
 \frac{x(u+i{Q \over g })-x(v+i{M \over g})}{x(u+i{Q \over g})-x(v-i{M \over g})}\,
 \frac{x(v-i{M \over g })}{x(v+i{M \over g})}~~~~
\\ &\times
&\prod_{j=1}^{Q-1}\frac{u-v-\frac{i}{g}(M-Q+2j)}{u-v+\frac{i}{g}(M-Q+2j)}\,, \label{Svwx}
\\\nonumber
\end{eqnarray}
and the dressing-phase related kernel $K_{{\mathfrak{sl}(2)}}^{QM}(u,v)$, which is built from the
${\mathfrak{sl}(2)}$ S-matrix of the model \cite{AFrev}. It is of the
form
\begin{eqnarray}
\label{Ssl2}
S_{{\mathfrak{sl}(2)}}^{QM}(u,v)= S^{QM}(u-v)^{-1} \,
\Sigma_{QM}(u,v)^{-2}\,,
\end{eqnarray}
where  $\Sigma^{QM}$ is the improved dressing factor \cite{dresscross}.
The corresponding ${\mathfrak{sl}(2)}$ and dressing kernels are defined in the
usual way
\begin{eqnarray}
K_{{\mathfrak{sl}(2)}}^{QM}(u,v)= \frac{1}{2\pi i}\frac{d}{du}\log
S_{{\mathfrak{sl}(2)}}^{QM}(u,v) \,,\quad K_{QM}^{\Sigma}(u,v)=\frac{1}{2\pi
i}\frac{d}{du}\log \Sigma_{QM}(u,v)\,.~~~~
\end{eqnarray}
Explicit expressions for the improved dressing factors $\Sigma_{QM}(u,v)$ 
can be found in section 6 of ref. \cite{dresscross}.
Here for our numerical computations we used the single integral representation
given in \cite{Frolov:2010wt}.

Finally we mention that along the lines of  \cite{BH2} 
in the derivation of the formula (\ref{OmegaKQ}) for $\Omega({\cal K}_Q)$, 
it was exploited that all the necessary kernels: \newline 
$K_{Q},K_{Qy},K^{Q1}_{xv},s \star K^{Q-1, 1}_{vwx},
K_{y1}, K^{Q 1}_{{\mathfrak{sl}(2)}}$ satisfy the identity:
\begin{equation}
{\cal K}_Q- s \star {\cal K}_{Q-1} -s \star {\cal K}_{Q+1}  \equiv\delta {\cal K}_Q=0, 
\qquad \mbox{for} \quad Q \geq 3. 
\end{equation}

\section{Kernel matrices of the vertical HNLIE part}

In this appendix the kernel matrices appearing in the upper HNLIE part of
our equations (\ref{bA},\ref{dA}) are presented. 
Here the kernel matrices are different compared to those published in \cite{BH1}.
The difference comes simply from a reformulation the equations in the language
of new unknown functions. In \cite{BH1} the unknowns are 6 $b$-type functions:
\begin{equation}
\underline{b}^{\scriptsize \mbox{old}}=\{ b^{(3)[\gamma_1]}_{1,s},b^{(3)[\gamma_2]}_{2,s},b^{(3)[\gamma_3]}_{3,s},
b^{(2)[-1+\gamma_4]}_{1,s},b^{(2)[-1+\gamma_5]}_{2,s},b^{(1)[-2+\gamma_6]}_{1,s} \},
\end{equation}
and 6 $d$-type functions:
\begin{equation}
\underline{d}^{\scriptsize  \mbox{old}}=\{ d^{(3)[\eta_1]}_{1,s},d^{(3)[\eta_2]}_{2,s},d^{(3)[\eta_3]}_{3,s},
d^{(2)[\eta_4]}_{1,s},d^{(2)[\eta_5]}_{2,s},d^{(1)[\eta_6]}_{1,s} \},
\end{equation}
with shift vectors $\underline{\gamma}$ and $\underline{\eta}$ given by (\ref{vgamma},\ref{veta}).
We recognized that the kernels become simpler if we formulate the equations in terms of
the unknowns:
\begin{equation}
\underline{b}=\{ b^{(3)[\gamma_1]}_{1,s},\eta/b^{(3)[\gamma_2]}_{2,s},b^{(3)[\gamma_3]}_{3,s},
\eta/b^{(2)[-1+\gamma_4]}_{1,s},b^{(2)[-1+\gamma_5]}_{2,s},b^{(1)[-2+\gamma_6]}_{1,s} \}, \label{buj}
\end{equation}
and
\begin{equation}
\underline{d}=\{ d^{(3)[\eta_1]}_{1,s},\eta/d^{(3)[\eta_2]}_{2,s},d^{(3)[\eta_3]}_{3,s},
d^{(2)[\eta_4]}_{1,s},\eta/d^{(2)[\eta_5]}_{2,s},d^{(1)[\eta_6]}_{1,s} \}, \label{duj}
\end{equation}
where $\eta=\pm 1$ is a global sign factor and $s=p_0$, if one adopts the notation of \cite{BH1}
for the HNLIE equations (\ref{bA},\ref{dA}). Another advantage of using the variables (\ref{buj},\ref{duj})
is that they are either $O(1)$ or exponentially small for large volumes.

For the sake of simplicity, here we give 
the form of the kernels of (\ref{bA},\ref{dA}) before the application of the contour shifts 
(\ref{vgamma},\ref{veta}). The kernels of the equations
can be obtained from these by simply shifting their arguments according to the
formulas below:
\begin{eqnarray}
G_{bB}(u)_{ab}&=&K_{bB}\left(u+\frac{i}{g} \, (\gamma_a-\gamma_b)\right)_{ab}, \qquad a,b=1,...,6 \nonumber \\
G_{bD}(u)_{ab}&=&K_{bD}\left(u+\frac{i}{g} \, (\gamma_a-\eta_b)\right)_{ab}, \qquad a,b=1,...,6 \nonumber \\
G_{dB}(u)_{ab}&=&K_{dB}\left(u+\frac{i}{g} \, (\eta_a-\gamma_b)\right)_{ab}, \qquad a,b=1,...,6 \nonumber \\
G_{dD}(u)_{ab}&=&K_{dD}\left(u+\frac{i}{g} \, (\eta_a-\eta_b)\right)_{ab}, \qquad a,b=1,...,6. \label{93a}
\end{eqnarray}
The kernel matrices can be expressed by the functions as follows\footnote{$\psi(z)=\frac{d}{dz} \log \Gamma(z)$.}:
\begin{eqnarray}
G(u)\!&=&\! \frac{g}{8 \, \pi} \left\{ \psi(1+\frac{i \, g \, u}{4})\! +\! \psi(1-\frac{i \, g \, u}{4})\!-\!
\psi(\frac12+\frac{i \, g \, u}{4})\!-\!\psi(\frac12+\frac{i \, g \, u}{4}) \right\},  \label{G}\\
l(u)\!&=&\!\frac{g}{8 \, \pi}  \left\{ \psi(1+\frac{i \, g \, u}{4})\! + \!\psi(1-\frac{i \, g \, u}{4}) \right\}, \label{l} \\
s(u)\!&=&\!\frac{g}{4} \, \frac{1}{\cosh(\frac{\pi g u}{2})} \label{s}
\end{eqnarray}
and they take the form:

{\begin{equation} \label{KbB}
K_{bB}=\left(
\begin{array}{cccccc}
G & 0 & G-s^+ & -s & s^+-G & 0 \\
0 & l & 0 & l-s^- & s & 0 \\
G-s^- & 0 & G & 0 & s^--G & 0 \\
s & l-s^+ & 0 & l & 0 & s  \\
s^--G & -s & s^+-G & 0 & G & s^-  \\
0 & 0 & 0 &-s & s^+ & G
\end{array}
\right),
\end{equation} }

{\begin{equation} \label{KbD}
K_{bD}=\left(
\begin{array}{cccccc}
-G & 0 & s^+-G & G-s^+ & 0 & 0 \\
0 & -l & 0 & -s & s^+-l & 0 \\
s^--G & 0 & -G & G-s^- & s & 0 \\
-s & s^+-l & 0 & 0 & s^+-l & s  \\
G-s^- & s & G-s^+ & -G & 0 & s^-  \\
0 & 0 & 0 & s^- & -s & -G^{--}
\end{array}
\right),
\end{equation} }

{\begin{equation} \label{KdB}
K_{dB}=\left(
\begin{array}{cccccc}
-G & 0 & s^+-G & s & G-s^+ & 0 \\
0 & -l & 0 & s^--l & -s & 0 \\
s^--G & 0 & -G & 0 & G-s^- & 0 \\
G-s^- & s & G-s^+ & 0 & -G & s^+  \\
0 & s^--l & -s & s^--l & 0 & s  \\
0 & 0 & 0 & -s & s^+ & -G^{++}
\end{array}
\right),
\end{equation} }

{\begin{equation} \label{KdD}
K_{dD}=\left(
\begin{array}{cccccc}
G & 0 & G-s^+ & s^+-G & 0 & 0 \\
0 & l & 0 & s & l-s^+ & 0 \\
G-s^- & 0 & G & s^--G & -s & 0 \\
s^--G & -s & s^+-G & G & 0 & s^+  \\
0 & l-s^- & s & 0 & l & s  \\
0 & 0 & 0 & s^- & -s & G
\end{array}
\right).
\end{equation} }

\section{Asymptotic solutions of the vertical HNLIE}

In this section along the lines of \cite{BH1} the asymptotic solutions of the upper $SU(4)$
NLIE variables are presented .
In the asymptotic limit the T-hook of AdS/CFT splits into two
$SU(2|2)$ fat-hooks.
The basic building blocks of the asymptotic solution are the
nine $Q$-functions corresponding to the left and right $SU(2|2)$ fat-hooks.
Due to the left-right symmetry of the $Y$-system it is enough to give
the right $Q$-functions.
They can be derived from the asymptotic solution of the Y-functions given in
\cite{DH}.
They take the form:
\begin{equation}
\begin{split}
Q^{(2,2)}(u)&=q_{22},\\
Q^{(2,1)}(u)&=\frac{2 \, q_{22}}{g\, \Lambda \, u^-},\\
Q^{(2,0)}(u)&=\frac{4 \,q_{22} \, u^{--}}{g\, u^- \, u^{---}}
\end{split}
\qquad\qquad
\begin{split}
Q^{(1,2)}(u)&=g\,\Lambda \, q_{11} \, \sigma(u) \, u^+,\\ 
Q^{(1,1)}(u)&= q_{11} \, \sigma(u),\\
Q^{(1,0)}(u)&=\Lambda \,q_{11} \, \sigma(u), 
\end{split}
\qquad\qquad
\begin{split}
Q^{(0,2)}(u)&= 4 \, g \, u^{++},\\
Q^{(0,1)}(u)&=\frac{2}{\Lambda},\\
Q^{(0,0)}(u)&=1,
\end{split}
\label{constantQ}
\end{equation}
where $q_{11},q_{22}$  and $\Lambda$ are arbitrary constants which
cancel from the final form of the asymptotic NLIE variables. 
Furthermore $\sigma(u)=e^{\frac{\pi \,g\, u}{2}}$ to satisfy
the recursion $\frac{\sigma^+}{\sigma^-}=-1$.
The further building blocks of the asymptotic solution are as follows:\footnote{
Here for correspondence we use the same letters for the names of different unknowns as in \cite{BH1}.}
\begin{equation}
T_{s,1}=\frac{4 \, (-1)^{s}\, s \, u} {u^{[s]}},
\end{equation}
and
\begin{equation}
{\cal A}^o(u)=\frac{4 \, u}{g\, u^+ \, u^-},\qquad {\cal B}^o(u)=4 \,g\, u, \qquad
\beta^o(u)=\frac{2 \, \sigma^-(u)}{\Lambda},\qquad \gamma^o(u)=\frac{2\, \sigma(u)}{g \, \Lambda \, u}.
\end{equation}
The solution of the recursions 
\begin{equation}
w^{o-}-w^{o+}=\frac{{\cal A}^o}{\gamma^{o+}\gamma^{o-}},\qquad\qquad
y^{o+}-y^{o-}=\frac{{\cal B}^o}{\beta^o\beta^{o--}},
\label{wyo}
\end{equation}
are as follows:
\begin{equation}
w^o(u)=-\frac{i \, \Lambda^2 \, e^{-\pi \, g\, u}}{4}((g \,u)^2+w_c),\qquad\qquad y^o(u)=\frac{i \, \Lambda^2 \, e^{-\pi \, g\, u}}{4}((g \,u)^2+w_c-i\, C),
\end{equation}
where $w_c$ and $C$ are arbitrary constants.
Using the building blocks listed above, the asymptotic form of the upper $SU(4)$ NLIE
functions can be determined \cite{BH1} and take the form:
\begin{eqnarray}
b^{(3)o}_{1,s}(u)&=&b^{(3)o}_{3,s}(u)=\frac{s \, u^+}{u^{[-s]}}, \label{bfirst}\\
B^{(3)o}_{1,s}(u)&=&B^{(3)o}_{3,s}(u)=\frac{(s+1) \, u}{u^{[-s]}},\\
b^{(3)o}_{2,s}(u)&=&B^{(3)o}_{2,s}(u)=-\phi^{[-s]}(u)\,\frac{1}{4\, g\, s\, u},\\
b^{(2)o-}_{1,s}(u)&=&B^{(2)o-}_{1,s}(u)=\frac{\phi^{[-s]}(u)}{4\, g\, s\, u+C},\\
b^{(2)o-}_{2,s}(u)&=&-\frac{C+4\, g\, s\, u^-}{4\, g\, u^{[-s]}},\qquad\qquad
B^{(2)o-}_{2,s}(u)=-\frac{C+4 \, g\, (s-1) \, u}{4\, g\, u^{[-s]}},\\
b^{(1)o--}_{1,s}(u)&=&\frac{C+4 \, g\, (s-1) \, u}{4\, g\, u^{[-s]}},\qquad\qquad
B^{(1)o--}_{1,s}(u)=\frac{C+4\, g\, s\, u^-}{4\, g\, u^{[-s]}},
\end{eqnarray}
\begin{eqnarray}
d^{(3)o}_{1,s}(u)&=&d^{(3)o}_{3,s}(u)=\frac{s \, u^-}{u^{[s]}},\\
D^{(3)o}_{1,s}(u)&=&D^{(3)o}_{3,s}(u)=\frac{(s+1) \, u}{u^{[s]}},\\
d^{(3)o}_{2,s}(u)&=&D^{(3)o}_{2,s}(u)=-\frac{1}{\phi^{[s]}(u)}\,\frac{(g\, u^{[s]})^2}{4\, g\, s\, u},\\
d^{(2)o}_{1,s}(u)&=&-\frac{C+4\, g\, s\, u^+}{4\, g\, u^{[s]}},\qquad\qquad
D^{(2)o}_{1,s}(u)=-\frac{C+4 \, g\, (s-1) \, u}{4\, g\, u^{[s]}},\\
d^{(2)o}_{2,s}(u)&=&D^{(2)o}_{2,s}(u)=-\frac{\phi^{[s]}(u) \, (g \,u^{[s]})^2}{4\, g\, s\, u+C},\\
d^{(1)o}_{1,s}(u)&=&\frac{C+4 \, g\, (s-1) \, u}{4\, g\, u^{[s]}},\qquad\qquad
D^{(1)o}_{1,s}(u)=\frac{C+4\, g\, s\, u^+}{4\, g\, u^{[s]}},
\end{eqnarray}
where $s$ is the "cutoff index" where the TBA $\rightarrow$ HNLIE replacements starts\footnote{In section 2.
it is denoted by $p_0$, here the notation $s$ is kept to fit to formulas of \cite{BH1}.}, furthermore
for any index distribution $B^o$ and $D^o$ stand for $1+b^o$ and $1+d^o$ respectively. 
\begin{equation}
\phi(u)=\frac{x(u)^{2L}}{g \,u},
\end{equation}
and $C$ is the arbitrary constant that does not cancel from the formula for the HNLIE variables.
The asymptotic solution for the six $b$- and $d$-type NLIE-functions of the system can be obtained
from the B\"acklund functions above by appropriately shifting their arguments:
\begin{equation}
{\bf b}^o=\{b^o_a\}=\{b^{(3)o[\gamma_1]}_{1,s},\eta/b^{(3)o[\gamma_2]}_{2,s},b^{(3)o[\gamma_3]}_{3,s},
\eta/b^{(2)o[-1+\gamma_4]}_{1,s},  b^{(2)o[-1+\gamma_5]}_{2,s}, b^{(1)o[-2+\gamma_6]}_{1,s}\}, \label{bvect}
\end{equation}
\begin{equation}
{\bf d}^o=\{d^o_a\}=\{d^{(3)o[\eta_1]}_{1,s},\eta/d^{(3)o[\eta_2]}_{2,s},d^{(3)o[\eta_3]}_{3,s},
d^{(2)o[\eta_4]}_{1,s},d^{(2)o[\eta_5]}_{2,s},d^{(1)o[\eta_6]}_{1,s}\}, \label{dvect}
\end{equation}
with the shifts given in (\ref{vgamma},\ref{veta}). Finally we note that the $C=0$ choice implies a symmetry relation between
the $b$- and $d$-type variables. Let ${\mathfrak M}$ the $6$ by $6$ matrix:
{\begin{equation} \label{M}
{\mathfrak M}=\left(
\begin{array}{cccccc}
0 & 0 & 1 & 0 & 0 & 0 \\
0 & 1 & 0 & 0 & 0 & 0 \\
1 & 0 & 0 & 0 & 0 & 0 \\
0 & 0 & 0 & 0 & 1 & 0  \\
0 & 0 & 0 & 1 & 0 & 0  \\
0 & 0 & 0 & 0 & 0 & 1
\end{array}
\right).
\end{equation} }
Then at $C=0$ the ${\bf b}^o$ and ${\bf d}^o$ vectors satisfy the relations as follows:
\begin{equation}
{\bf d}^o(u)={\mathfrak M}{\bf b}^o(-u), \qquad {\bf b}^o(-u)={\bf b}^{o*}(u), \label{MbO}
\end{equation}
\begin{equation}
{\bf d}^{o}(u)={\mathfrak M}{\bf b}^{o*}(u), \qquad {\bf d}^o(-u)={\bf d}^{o*}(u),
\end{equation}
where $*$ denotes complex conjugation. In our numerical studies we mostly use the $C=0$
asymptotic solution to setup the equations to solve. In this case the exact equations
guarantee the fulfillment of (\ref{bmMd}), which reduces to $6$ the number of independent
complex functions of the upper NLIE part.

\section{Asymptotic solutions of the $Y$-system and the horizontal $SU(2)$-type HNLIE}

This appendix is devoted to give the asymptotic solution for the $Y$-functions and the
variables  of the horizontal $SU(2)$ NLIE. The asymptotic form of the $Y$-functions
can be read off from the asymptotic $T$-functions in \cite{DH}. They take the form:
\begin{equation}
Y_{m|vw}^o(u)=\frac{m(m+2)\,g^2 \, u^2}{(m+1)^2+g^2 \, u^2}, \quad m=1,2,...
\end{equation}
\begin{equation}
Y_Q^o(u)=\left(\frac{1}{x_s^{[Q]}(u)\, x_s^{[-Q]}(u)}\right)^{2L} \frac{16 \, Q^2 \, g^2 \, u^2}{g^2\, u^2 +Q^2}, 
\qquad Q=1,2,...\label{YQo}
\end{equation}
\begin{equation}
Y_-^o(u)=Y_+^{o}(u)=-\frac{g^2 \, u^2}{2+g^2 \, u^2},
\end{equation}
\begin{equation}
Y_{1|w}^o(u)=\frac{g^2 \, u^2 \, (19+3 \, g^2 \, u^2)}{(1+g^2 u^2)(4+g^2 u^2)}.
\end{equation}
Following the lines of \cite{BH1} the asymptotic horizontal $SU(2)$ NLIE variables can be determined 
from the asymptotic $Q$-functions (\ref{constantQ}). 
Here we just list the final formulas:
\begin{equation}
b^o(u)=b_0(u-i\, \gamma), \qquad \bar{b}^o(u)=\bar{b}_0(u+i\, \gamma),
\end{equation}
where 
\begin{equation}
b_0(u)=\frac{2 \, (g^2 \, u^2-3 \, i)\,(1+2 \, i\, g \, u+g^2 \, u^2)}{(g^2 \, u^2+i)\, (g^2 \, u^2-2 \, i) \, (g^2 \, u^2+3\, i)},
\end{equation}
\begin{equation}
\bar{b}_0(u)=\frac{2 \, (g^2 \, u^2+3 \, i)\,(1-2 \, i\, g \, u+g^2 \, u^2)}{(g^2 \, u^2-i)\, (g^2 \, u^2+2 \, i) \, (g^2 \, u^2-3\, i)},
\end{equation}
and $0<\gamma<1/2$ is the arbitrary contour shift parameter of the horizontal $SU(2)$ NLIE.


\newpage


\begin{thebibliography}{99}
\bibitem{adscft1} J.~M.~Maldacena,
{\it{``The large N limit of superconformal field theories and supergravity,''}}
{\em Adv.\ Theor.\ Math.\ Phys.\ }  {\bf 2} (1998) 231, 
[{\em Int.\ J.\ Theor.\ Phys.\ }  {\bf 38} (1999) 1113];
\bibitem{adscft2}
S.~S.~Gubser, I.~R.~Klebanov and A.~M.~Polyakov,
{\it {``Gauge theory correlators from non-critical string theory,''}}
{\em Phys.\ Lett.\ B} {\bf 428} (1998) 105;
\bibitem{adscft3}
E.~Witten,
{\it{``Anti-de Sitter space and holography,''}}
{\em Adv.\ Theor.\ Math.\ Phys.\ }  {\bf 2} (1998) 253;

\bibitem{DH}
Z. Bajnok, N. Drukker, \'A. Heged\H{u}s, R. I. Nepomechie, L. Palla, C. Sieg,
R. Suzuki,
{\it {"The spectrum of tachyons in AdS/CFT"}}
{\em JHEP} {\bf 1403} (2014) 055.

\bibitem{Over}
  N.~Beisert, C.~Ahn, L.~F.~Alday, Z.~Bajnok, J.~M.~Drummond, L.~Freyhult, N.~Gromov and R.~A.~Janik {\it et al.},
  {\it {``Review of AdS/CFT Integrability: An Overview,''}}
 {\em Lett.\ Math.\ Phys.\ } {\bf 99} (2012) 3.

\bibitem{DeWolfe:2004zt}
O.~DeWolfe and N.~Mann, {\it {Integrable open spin chains in defect conformal
  field theory}},  {\em JHEP} {\bf 0404} (2004) 035,

\bibitem{Berenstein:2005vf}
D.~Berenstein and S.~E. Vazquez, {\it {Integrable open spin chains from giant
  gravitons}},  {\em JHEP} {\bf 0506} (2005) 059,

\bibitem{Hofman:2007xp}
D.~M. Hofman and J.~M. Maldacena, {\it {Reflecting magnons}},  {\em JHEP} {\bf
  0711} (2007) 063,

\bibitem{Correa:2008av}
D.~Correa and C.~Young, {\it {Reflecting magnons from D7 and D5 branes}},  {\em
  J.Phys.A} {\bf A41} (2008) 455401,

\bibitem{CRY}
D.~H. Correa, V.~Regelskis, and C.~A. Young, {\it {Integrable achiral D5-brane
  reflections and asymptotic Bethe equations}},  {\em J.Phys.A} {\bf A44}
  (2011) 325403,



\bibitem{Bajnok:2012bz}
  Z.~Bajnok and R.~A.~Janik,
  {\it {``Six and seven loop Konishi from Luscher corrections,''}}
  {\em JHEP} {\bf 1211} (2012) 002

 \bibitem{Leurent:2012ab}
  S.~Leurent, D.~Serban and D.~Volin,
  {\it{``Six-loop Konishi anomalous dimension from the Y-system,''}}
  Phys.\ Rev.\ Lett.\  {\bf 109} (2012) 241601

\bibitem{Leurent:2013mr}
  S.~Leurent and D.~Volin,
  {\it{``Multiple zeta functions and double wrapping in planar $N=4$ SYM,''}}
  Nucl.\ Phys.\ B {\bf 875} (2013) 757


\bibitem{Volinuj}
C.~Marboe, D.~Volin, {\it{"Quantum spectral curve as a tool
 for a perturbative quantum field theory" }}
\href{http://arxiv.org/abs/1411.4758} {\path{arXiv:1411.4758}}.

\bibitem{Marboe:2014sya}
  C.~Marboe, V.~Velizhanin and D.~Volin,
  {\it{``Six-loop anomalous dimension of twist-two operators in planar N=4 SYM theory,''}}
  arXiv:1412.4762 [hep-th].

 \bibitem{Gromov:2009tq}
  N.~Gromov,
  {\it{``Y-system and Quasi-Classical Strings,''}}
  JHEP {\bf 1001} (2010) 112.
 
 \bibitem{Gromov:2010vb}
  N.~Gromov, V.~Kazakov and Z.~Tsuboi,
  {\it{``$PSU(2,2|4)$ Character of Quasiclassical AdS/CFT,''}}
  JHEP {\bf 1007} (2010) 097


 \bibitem{Bajnok:2013sza}
  Z.~Bajnok, M.~Kim and L.~Palla,
 {\it {``Spectral curve for open strings attached to the Y=0 brane,''}}
  {\em JHEP} {\bf 1404} (2014) 035

\bibitem{Gromov:2014bva}
  N.~Gromov, F.~Levkovich-Maslyuk, G.~Sizov and S.~Valatka,
  {\it {``Quantum spectral curve at work: from small spin to strong coupling in $ \mathcal{N} $ = 4 SYM,''}}
  JHEP {\bf 1407} (2014) 156

\bibitem{Gromov:2009zb}
  N.~Gromov, V.~Kazakov and P.~Vieira,
  {\it {``Exact Spectrum of Planar ${\cal N}=4$ Supersymmetric Yang-Mills Theory: Konishi Dimension at Any Coupling,''}}
  Phys.\ Rev.\ Lett.\  {\bf 104} (2010) 211601


\bibitem{Frolov:2010wt}
S.~Frolov, {\it {Konishi operator at intermediate coupling}},  {\em J.Phys.}
  {\bf A44} (2011) 065401,

\bibitem{Frolov:2012zv}
  S.~Frolov,
 {\it {``Scaling dimensions from the mirror TBA,''}}
  J.\ Phys.\ A {\bf 45} (2012) 305402.



\bibitem{McGreevy:2000cw}
J.~McGreevy, L.~Susskind, and N.~Toumbas, {\it {Invasion of the giant gravitons
  from Anti-de Sitter space}},  {\em JHEP} {\bf 0006} (2000) 008,

\bibitem{Balasubramanian:2001nh}
V.~Balasubramanian, M.~Berkooz, A.~Naqvi, and M.~J. Strassler, {\it {Giant
  gravitons in conformal field theory}},  {\em JHEP} {\bf 0204} (2002) 034,

\bibitem{DH33}
V.~Balasubramanian, M.-x. Huang, T.~S. Levi, and A.~Naqvi, {\it {Open strings
  from ${\cal N}=4$ superYang-Mills}},  {\em JHEP} {\bf 0208} (2002) 037,
 
 
\bibitem{LeClair:1995uf}
A.~LeClair, G.~Mussardo, H.~Saleur, and S.~Skorik, {\it {Boundary energy and
  boundary states in integrable quantum field theories}},  {\em Nucl.Phys.}
  {\bf B453} (1995) 581--618,


\bibitem{Correa:2009mz}
D.~Correa and C.~Young, {\it {Finite size corrections for open strings/open
  chains in planar $AdS$/CFT}},  {\em JHEP} {\bf 0908} (2009) 097,

\bibitem{Palla:2011eu}
L.~Palla, {\it {Yangian symmetry of boundary scattering in AdS/CFT and the
  explicit form of bound state reflection matrices}},  {\em JHEP} {\bf 1103}
  (2011) 110, 

\bibitem{Ahn:2010xa}
C.~Ahn and R.~I. Nepomechie, {\it {Yangian symmetry and bound states in AdS/CFT
  boundary scattering}},  {\em JHEP} {\bf 1005} (2010) 016,

\bibitem{Dru-int-WL}
N.~Drukker, {\it {Integrable Wilson loops}},
{\em JHEP} {\bf 1310} (2013) 135,

\bibitem{CMS}
D.~Correa, J.~Maldacena, and A.~Sever, {\it {The quark anti-quark potential and
  the cusp anomalous dimension from a TBA equation}},
  {\em JHEP} {\bf 1208} (2012) 134,

\bibitem{Bajnok:2013sya}
  Z.~Bajnok, J.~Balog, D.~H.~Correa, \'A.~Heged\H{u}s, F.~I.~Schaposnik Massolo and G.~Zsolt T\'oth,
  {\it {``Reformulating the TBA equations for the quark anti-quark potential and their two loop expansion,''}}
  {\em JHEP} {\bf 1403} (2014) 056,


\bibitem{GKV09}
  N.~Gromov, V.~Kazakov and P.~Vieira,
  {\it {``Integrability for the Full Spectrum of Planar AdS/CFT,''}}
{\em  Phys.Rev.Lett. }{\bf 103:131601},2009,

\bibitem{Bombardelli:2009ns}
  D.~Bombardelli, D.~Fioravanti and R.~Tateo,
 {\it {``Thermodynamic Bethe Ansatz for planar AdS/CFT: A Proposal,''}}
  J.\ Phys.\ A {\bf 42} (2009) 375401

\bibitem{Arutyunov:2009ur}
  G.~Arutyunov and S.~Frolov,
 {\it { ``Thermodynamic Bethe Ansatz for the $AdS_5 \times S^5$ Mirror Model,''}}
  JHEP {\bf 0905} (2009) 068

\bibitem{Cavaglia:2010nm}
A.~Cavaglia, D.~Fioravanti, and R.~Tateo, {\it {Extended Y-system for the
  $AdS_5/CFT_4$ correspondence}},  {\em Nucl.Phys.} {\bf B843} (2011) 302--343,

 \bibitem{Balog:2011nm}
  J.~Balog and A.~Hegedus,
 {\it {``$AdS_5\times S^5$ mirror TBA equations from Y-system and discontinuity relations,''}}
  JHEP {\bf 1108} (2011) 095
 
\bibitem{BH1}
J. Balog, \'A. Heged\H{u}s, {\it {"Hybrid-NLIE for the AdS/CFT spectral problem"}}
{\em JHEP} {\bf 1208} (2012) 022.

\bibitem{Bajnok:2010ui}
Z.~Bajnok and L.~Palla, {\it {Boundary finite size corrections for
  multiparticle states and planar $AdS$/CFT}},  {\em JHEP} {\bf 1101} (2011)
  011,

\bibitem{AFshyp}
G. Arutyunov, S. Frolov, R. Suzuki, {\it {"String hypothesis for the $AdS_5 \times S^5$ mirror"}}
{\em JHEP}{\bf 03} (2009) 152.

\bibitem{AFS}
G. Arutyunov, S. Frolov, R. Suzuki, {\it {"Exploring the mirror TBA"}}
{\em JHEP} {\bf 05} (2010) 031.

\bibitem{BH2}
J. Balog, \'A. Heged\H{u}s, { \it {"Quasi-local formulation of the mirror TBA" }}
{\em JHEP} {\bf 1205} (2012) 039.

\bibitem{RyoNLIE}
 R. Suzuki, {\it {"Hybrid NLIE for the mirror  $AdS_5 \times S^5$ "}}
{\em J.Phys.}{\bf A44} (2011) 235401.

\bibitem{Destri:1997yz}
  C.~Destri and H.~J.~de Vega,
  {\it {``Nonlinear integral equation and excited states scaling functions in the sine-Gordon model,''}}
 {\em Nucl.\ Phys.\ B} {\bf 504} (1997) 621-644

\bibitem{Hegedus:2007jw}
  A.~Hegedus,
  {\it {``Finite size effects and 2-string deviations in the spin-1 XXZ chains,''}}
  {\em J.\ Phys.\ A} {\bf 40} (2007) 12007

 \bibitem{Gromov:2011cx}
  N.~Gromov, V.~Kazakov, S.~Leurent and D.~Volin,
  {\it {``Solving the AdS/CFT Y-system,''}}
  JHEP {\bf 1207} (2012) 023
 
\bibitem{Gromov:2013pga}
  N.~Gromov, V.~Kazakov, S.~Leurent and D.~Volin,
  {\it {``Quantum Spectral Curve for Planar $\mathcal{N} =$ Super-Yang-Mills Theory,''}}
  Phys.\ Rev.\ Lett.\  {\bf 112} (2014) 1,  011602


\bibitem{Gromov:2014caa}
N.~Gromov, V.~Kazakov, S.~Leurent, D.~Volin, {\it{Quantum spectral curve for
  arbitrary state/operator in AdS$_5$/CFT$_4$}}
\href{http://arxiv.org/abs/1405.4857} {\path{arXiv:1405.4857}}.




  
  

\bibitem{AFrev}
  G.~Arutyunov and S.~Frolov,
  {\it{``Foundations of the $AdS_5 \times S^5$ Superstring. Part I,''}}
  J.\ Phys.\ A  {\bf 42} (2009) 254003


\bibitem{dresscross}
G. Arutyunov, S. Frolov, {\it{``The dressing factor and crossing
equations''}}, J.\ Phys.\ A  {\bf 42}, 425401 (2009) 






 
 
 
 
\end{thebibliography}
\end{document}